\documentclass[aps,prd,reprint,amsmath,amssymb,nobibnotes,showpacs]{revtex4-1}

\usepackage{graphicx}
\usepackage{dcolumn}
\usepackage{amsfonts}
\usepackage{bm}
\begin{document}
\title{Exploring the effects of a double reconstruction on the geometrical parameters of coupled models, using observational data}
\author{Freddy Cueva Solano}
\affiliation{Instituto de F\'{\i}sica y Matem\'aticas, Universidad Michoacana de San Nicol\'as de Hidalgo\\
Edificio C-3, Ciudad Universitaria, CP. 58040, Morelia, Michoac\'an, M\'exico.}
\email{freddy@ifm.umich.mx,\;\;freddycuevasolano$2009$@gmail.com}
\date{\today}
\begin{abstract}
In this work we study the effects of the non-gravitational exchange energy ($Q$) between dark matter ($DM$) fluid and dark energy ($DE$) fluid on the background 
evolution of the cosmological parameters. A varying equation of state (EOS) parameter, $\omega$, for $DE$ is proposed. Considering an 
universe spatially flat, two distinct coupled models were examined to explore the main cosmological effects generated by the simultaneous reconstruction of 
$Q$ and $\omega$ on the shape of the jerk parameter, $j$, through a slight enhancement or suppression of their amplitudes with respect to 
uncoupled scenarios, during its evolution from the past to the near future. In consequence, $j$ could be used to distinguish any coupled $DE$ models.
Otherwise, the observational data were used to put stringent constraints on $Q$ and $\omega$, respectively. In such a way, we used our results as evidences to search 
possible deviations from the standard concordance model ($\Lambda$CDM), examining their predictions and improving our knowledge of the cosmic evolution of the 
universe.
\end{abstract}
\pacs{98.80.-k, 95.35.+d, 95.36.+x, 98.80.Es}
\maketitle
\section{Introduction}
The recent astronomical measurements of type Ia Supernovae Union $2.1$ (Union $2.1$ SNIa) composed by $580$ data \cite{Riess1998,AmanullahUnion22010,nesseris1,
Suzuki2012}, the Baryon Acoustic Oscillation (BAO) detected in the clustering of the combined $2$dF Galaxy Survey ($2$dFGRS) and the Sloan Digital 
Sky Survey (SDSS) Data Release $7$ (DR $7$) main galaxy samples, the $6$dF Galaxy Survey ($6$dFGS) and the WiggleZ Dark Energy Survey (WiggleZ)
\cite{Eisenstein1998,SDSS,ReidSDSS,Beutler2011,Blake2011}, the observations of polarization and anisotropies in the power 
spectrum of the Cosmic Microwave Background (CMB: distance priors) data from the Wilkinson Microwave Anisotropy Probe $7$ year (WMAP $7$)
\cite{Hu-Sugiyama1996,Bond-Tegmark1997,WMAP,Komatsu2011}, the observational Hubble (H) data set measured from galaxy surveys 
\cite{Jimenez2002,Jimenez2003,Simon2005,Hubble2009-Stern2010,Gaztanaga}, 
and other, have confirmed that the present universe is undergoing an accelerated phase of expansion.
In the literature, some theoretical approaches were taken into account to explain this phenomenon, we are interested in an universe in where exists 
an exotic energy component with negative pressure, named $DE$ \cite{Peebles1988,Peebles2003,Sahni2004,Copeland2006}, and which presumably began to dominate 
the evolution of the universe, only recently. Within this approach, the simplest candidate for $DE$ is the Cosmological Constant $\Lambda$, which has an EOS parameter
$\omega=-1.0$ \cite{Weinberg1989,Sahni2000,Seljak2005,Rozo2010}. Also, there exist other alternative models such as phantom model \cite{Caldwell2002}, quintom model 
\cite{Feng2005}, quintessence model \cite{Ratra1988}, the k-essence model \cite{Picon-Chiba}, Chaplyging gas model \cite{Pasquier-Harko}, 
massive scalar field model \cite{Garousi-Sami} and other. All these models predict different dynamics of the universe.\\ 
On the other hand, the properties of $DE$ are mainly cha-racterized by $\omega$. In such a way, due to our ignorance of its nature, it was parameterize empirically
in a model independent way. In this sense, we have followed two ways to explore its behaviour. The first one was to paramete-rize $\omega$ 
in terms of some free parameters \cite{Cooray1999,Chevallier-Linder,Tegmark2004,Barboza2009,Wu2010}. 
Among all the different parametrizations forms the Chevallier-Polarski-Linder (CPL) parametrization \cite{Chevallier-Linder} is  considered as the most popular ansatz 
$\omega=\omega_{o}+\omega_{1}[z/(1+z)]$, where $z$ is the redshift and $\omega_{o}$, $\omega_{1}$ are real parameters \cite{Chevallier-Linder}. This ansatz 
has a divergence problem, when redshift $z$ approaches to $-1$ \cite{Li-Ma}. In addition, some non-parametric forms were found in \cite{Holsclaws}.
The second one was to choose an appropiated local basis representation for $\omega$ and after estimate the associated coefficients 
\cite{Daly2003,Huterer2005,Alam2006,Hojjati2010}. 
However, a divergence-free reconstruction for $\omega$ was proposed here, expanding $\omega$ in terms of the Chebyshev 
polynomials $T_{n}, n \in N$. To display how the method runs $\omega$ was expanded in terms of only the first 
three Chebyshev polynomials $T_{n},\,n=0,1,2$, and therefore, they are considered as a complete orthonormal basis on the finite interval 
$[-1,1]$, and besides, belong to the Hilbert space $L^{2}$ of real values \cite{Olivier2012}. They were chosen because have the property to be 
the minimal approximately polynomials \cite{Simon2005, Martinez2008}.\\
On the other hand, within the universe another dark component has been assumed its existence, so-called $DM$, 
which acts exactly like the ordinary matter (pre-ssureless), but does not interact with $DE$, except gravi-tationally. 
The nature of these dark components are not still known and the possibility that within the universe 
exists a non-gravitational coupling in the dark sector could not be precluded, as well as, its possible effect on the dynamics evolution of the cosmological parameters
should be considered \cite{Turner1983,Malik2003,Cen2003,Guo2007,Bohmer2008,valiviita2008,campo2009,cabral2009,chimento2010,abramo1,Cai-Su,abramo2,cao2011,LiZhang2011}. 
Some consequences of it were already studied in \cite{Zimdahl2005,Das2006,Huey2006,Wang2007}, from which the strenght of the coupling should be very small.\\
A huge amount of coupled $DE$ models have already been investigated and fitted with cosmological data. Some of them were 
motivated by mathematical simplicity, for example, models in which $Q\propto H\rho$, in where $H$ and $\rho$ denote the Hubble parameter and the energy density
of dark sectors, respectively. It has three possibilities, namely, $\rho=\rho_{DM}$ ($DM$ energy density), $\rho=\rho_{DE}$ ($DE$ energy density) and
$\rho=\rho_{DM}+\rho_{DE}$ \cite{LiZhang2011}. On the contrary, the models with $Q \propto \rho$ have been used in rehe-ating 
\cite{Turner1983}, curvaton decay \cite{Malik2003} and decay of $DM$ into radiation \cite{Cen2003}. All these models strongly 
depend on the choice made for the $Q$ form. So far, the coupled $DE$ models have not been investigated, in a general form before. In effect, some attempt 
to reconstruct $Q$ from a general parametrization has been done by us in \cite{Cueva-Nucamendi2012}.
In this paper we have considered different theoretical scenarios, in where $\omega$ was always taken as constant. Here, we fixed
$Q\propto H\rm {I}_{Q}(1+z)^{3}$, where the function $\rm {I}_{Q}$ was reconstructed in terms of the first six Chebyshev polynomials. 
The analysis was done using a sample of SNIa Union $2$ data. Our main results showed that the best fitted on $\rm {I}_{Q}$ have prefered to cross 
the noncoupling line $\rm {I}_{Q}=0$ during its evolution.\\
The motivation of this article has been to go from theory to observations, following the prescription outlines by Cueva-Nucamendi \cite{Cueva-Nucamendi2012}.
Then, to follow this thread, a coupled model with two reconstructions describing to $Q$ and $\omega$ simultaneously has been proposed here. Therefore, we have postulated 
the existence of a general non-gravitational coupling between $DE$ and $DM$ \cite{Cueva-Nucamendi2012}, introducing a general phenomenological parametrization for $Q$ 
into the equations of motion of these dark components. Here, $Q$ was reconstructed expanding it in terms of the first three Chebyshev polynomials 
$T_{n},\,n=0,1,2$. This has been the first attempt at reconstructing simultaneously $Q$ and $\omega$ from real data.\\
Two distinct coupled $DE$ models such as XCPL and DR were analysed here. Within these scenarios, the aim of our paper has been to study the effects that result from
the reconstructions of $Q$ and $\omega$ on the cosmological background evolution, of some parameters such as (defined below) $DM$ energy density parameter ($\Omega_{DM}$), 
deceleration parameter ($q$) and jerk parameter ($j$) \cite{Qi2009,Rubano2012,Viel2010,Gruber2012,Sendra2013,Gruber2014}, whose amplitudes are modified with respect to 
those of uncoupled models. Here, our models were constrained using an analysis combined of Union $2.1$ SNIa \cite{Riess1998,AmanullahUnion22010,nesseris1,Suzuki2012}, 
BAO \cite{Eisenstein1998,SDSS,ReidSDSS,Beutler2011,Blake2011}, CMB \cite{Hu-Sugiyama1996,Bond-Tegmark1997,WMAP,Komatsu2011}, and H data 
sets \cite{Jimenez2002,Jimenez2003,Simon2005,Hubble2009-Stern2010,Gaztanaga}.\\
Finally, we organize this paper as follows: The background equation of motions for the energy densities, the definition of the geometrical parameters, and the 
reconstruction schemes for $Q$ and $\omega$ are derived in section II.\\
In section III we describe the coupled $DE$ models worked. The priors considered and the observational 
constraints on the parameters space are discussed in section IV. We discuss our results in section V. In section VI we conclude our main results.
\section{Background equations of motion}\label{Background}
In a flat Friedmann-Robertson-Walker (FRW) universe its background dynamics is described by the following set of equations for their energy densities
(detailed calculations are found in \cite{Cueva-Nucamendi2012}, so we do not discuss them here.)
\begin{eqnarray}
\label{EoFB}{\dot{{\rho}}}_{b}+3{H}\;{{\rho}}_{b}&=&0\;,\\
\label{EoFr}{\dot{{\rho}}}_{r}+4{H}\;{{\rho}}_{r}&=&0\;,\\
\label{EoFDM}{\dot{{\rho}}}_{DM}+3{H}\;{{\rho}}_{DM}&=&+{Q}\;,\\
\label{EoFDE}{\dot{{\rho}}}_{DE}+3 \left(1+{\omega}\right)\;{H}\;{{\rho}}_{DE}&=&-{Q}\;,
\end{eqnarray}
where $\rho_{b}$, $\rho_{r}$, $\rho_{DM}$ and $\rho_{DE}$ are the energy densities of the baryon, radiation, $DM$ and $DE$, respectively. Now defined the Hubble 
expansion rate as $H \equiv \dot{a}/a$, and also, ``$\cdot$'' indicates differentiation with respect to the time $t$.\\
In what follows we shall assume that there is not ener-gy transfer from $DE$ ($DM$) to baryon or radiation, and among them only exist a gravitational coupling 
\cite{Koyama2009-Brax2010}.  
The critical densities $\rho_{c}\equiv3H^2/8\pi G$, and the critical density today $\rho_{c,0}\equiv 3H^2_{0}/8\pi G$, in where
$H_{0}$ is the current value of the Hubble parameter, were conveniently defined. Considering that $A=b, r, DM, DE$, then the normalized densities are
\begin{equation}\label{energydensity} 
{\Omega}_{A}\equiv\frac{{\rho}_{A}}{\rho_{c}}=\frac{{\rho}_{A}/\rho_{c,0}}{\rho_{c}/\rho_{c,0}}=\frac{\Omega^{\star}_{A}}{E^{2}}\;,\qquad{\Omega}_{A,0}\equiv
\frac{{\rho}_{A,0}}{\rho_{c,0}}\;.
\end{equation}
The first Friedmann equation is then given by
\begin{eqnarray}\label{hubble} 
E^{2}\equiv\frac{{H}^{2}}{{H}^{2}_{o}}&=&\frac{8\pi G}{3{H}^{2}_{o}}\left({\rho}_{b}+{\rho}_{r}+{\rho}_{DM}+{\rho}_{DE}\right)\;,\nonumber\\
&&=\left[{\Omega}^{\star}_{b}+{\Omega}^{\star}_{r}+{\Omega}^{\star}_{DM}+{\Omega}^{\star}_{DE}\right]\;,
\end{eqnarray}
and with the following relation for all time
\begin{equation}\label{OmegaDE_present} 
\Omega_{b}+\Omega_{r}+\Omega_{DM}+\Omega_{DE}=1\;\;.  
\end{equation}
The scale factor $a$ is related with the redshift through $a=1/(1+z)$, from which find $dt/dz=-1/(1+z)H(z)$. By substituting this last relation into
Eqs.\;(\ref{EoFB})-(\ref{EoFDE}), and solving Eqs.\;(\ref{EoFB})-(\ref{EoFr}), find the redshift evolution of $\Omega^{\star}_{A}$
\begin{eqnarray}
\label{Omegabs} \Omega^{\star}_{b}(z)&=&\Omega_{b,0}{(1+z)}^{3}\;,\\
\label{Omegars} \Omega^{\star}_{r}(z)&=&\Omega_{r,0}{(1+z)}^{4}\;,\\
\label{EDFDMOmega} 
\frac{\mathrm{d}\Omega^{\star}_{DM}}{\mathrm{d}z}-\frac{3\,\Omega^{\star}_{DM}}{1+z}&=&\frac{-\Omega^{\star}_{DM}{\rm I}_{\rm Q}(z)}{1+z}\,,\quad\\
\label{EDFDEOmega} 
\frac{\mathrm{d}\Omega^{\star}_{DE}}{\mathrm{d}z}-\frac{3(1+\omega)\,\Omega^{\star}_{DE}}{1+z}&=&\frac{+\Omega^{\star}_{DM}{\rm I}_{\rm Q}(z)}{1+z}\,.
\end{eqnarray}
Then, these equations have been fundamental to determine the results within our models.
\subsection{Evolution of geometrical parameters}\label{geometrical}
The geometrical parameters of the universe are obtained by performing a Taylor series expansion of the scale factor $a(t)$ around the current epoch, $t_{0}$.
Conventionally, this series is truncated at a determined order \cite{Qi2009}. Then, in this work we have been truncated such series 
at third order to study its behaviour, in where the dimensionless coefficients such as deceleration parameter, $q$, and jerk parameter, $j$, are defined 
as \cite{Qi2009}:
\begin{eqnarray}
\label{deceleration_definition} 
q(z)\equiv-\frac{1}{H^{2}}\frac{\ddot{a}}{a}&=&-1+\frac{(1+z)}{H(z)}H'(z)\;,\\
\label{jerk_definition}
j(z)\equiv +\frac{1}{H^{3}}\frac{\dddot{a}}{a}&=&q(z)+2q^{2}(z)+(1+z)q'(z),
\end{eqnarray}
where $\ddot{a}$ and $\dddot{a}$ are the second and third derivatives of $a$ with respect to time, respectively, and also, 
``\;$'$\;'' indicates differentiation with respect to $z$. Besides, some authors define the jerk parameter, $j$, with opposite sign 
\cite{Qi2009, Viel2010, Gruber2014}. Let us substitute Eqs. (\ref{energydensity})-(\ref{OmegaDE_present}) into Eq. (\ref{deceleration_definition}), 
\begin{equation}\label{q_values}
q(z)=+\frac{3}{2}\omega(z)\Omega_{DE}(z)+\frac{1}{2}+\displaystyle{\frac{\Omega_{r}(z)}{2}},
\end{equation}
and its derivative is 
\begin{equation}\label{q'_values}
q'(z)=+\frac{3}{2}\omega'(z)\Omega_{DE}(z)+\frac{3}{2}\omega(z)\Omega_{DE}'(z)+\displaystyle{\frac{\Omega'_{r}(z)}{2}}\;.
\end{equation}
These equations are frequently used in this work.
\subsection{Parametrizations of $Q$ and $w$} \label{General Reconstruction}
The Chebyshev polynomials form a complete set of orthonormal functions on the interval $[-1, 1]$ and have the property to be the minimal approximating polynomials, 
which means that have the smallest maximum deviation from the true function at any given order \cite{Simon2005, Cueva-Nucamendi2012}.\\ 
In general, an energy exchange is described as the coupling between both dak fluids and phenomenologically it is chosen as a rate proportional to $H$
\begin{equation}\label{Interaction} 
{{Q}}\equiv H\rho_{DM}{\rm I}_{\rm Q}\;.
\end{equation}
Here, the strength of the coupling is characterized by ${{\rm{I}}}_{\rm{Q}}$,
\begin{equation}\label{CouplingIq} 
{{\rm I}}_{\rm Q}\equiv \sum_{n=0}^{2}{{\lambda}}_{n}T_{n}\;,
\end{equation}
in where the coefficients of the polynomial expansion ${\lambda}_{n}$ are free dimensionless parameters \cite{Cueva-Nucamendi2012} and
\begin{equation}\label{Chebyshev1} 
T_{0}(z)=1\;,\hspace{0.3cm}T_{1}(z)=z\;,\hspace{0.3cm} T_{2}(z)=(2z^{2}-1)\;, 
\end{equation}
represent the first three Chebyshev polynomials.\\
Within the CPL model, the past evolution history may be successfully described by its EOS parameter, $\omega$, but the future evolution may not be explained,
because $\omega$ grows increasingly, and then, encounters a divergence when $z\rightarrow -1$. This is not a physical feature. Therefore, a novel reconstruction
form for $\omega$ has been proposed here to avoid such divergence problem. Hence, $\omega$ is defined as  
\begin{equation}\label{CouplingIw} 
{{\omega}}\equiv \sum_{m=0}^{2}{{\omega}}_{m}T_{m}\;,
\end{equation}
where $\omega_{o}, \omega_{1}$ and $\omega_{2}$ are free dimensionless parameters.\\ 
The Chebyshev polynomials of order $m=2$ were defined by Eq. (\ref{Chebyshev1}). Using numerical simulations we will compute the best fitted values for 
${\lambda}_{0}$, ${\lambda}_{1}$, ${\lambda}_{2}$, $\omega_{o}$, $\omega_{1}$ and $\omega_{2}$, respectively.
\section{Dark\,energy\,models} \label{DE_models}
 \subsection{$\Lambda$CDM model}\label{LCDM}
In this scenario, the function $E^{2}$ was found fixing both $\omega(z)=-1$ and $Q(z)=0$ into Eqs. (\ref{Omegabs})-(\ref{EDFDEOmega})
\begin{equation}\label{hubble_LCDM}
E^{2}=\biggl[\Omega^{\star}_{b}(z)+\Omega^{\star}_{r}(z)+\Omega_{DM,0}{(1+z)}^{3}+\Omega_{DE,0}\biggr]\;,
\end{equation}
moreover, $q$ and $q'$ are given by the Eqs. (\ref{q_values})-(\ref{q'_values}), then
\begin{eqnarray}\label{q_lcdm}
q(z)&=&+\frac{3}{2}\biggl[\Omega_{b}(z)+\Omega_{DM}(z)\biggr]+2\,{\Omega_{r}(z)}-1\;,\nonumber\\
q'(z)&=&+\frac{3}{2}\biggl[\Omega_{b}'(z)+\Omega_{DM}'(z)\biggr]+2\,{\Omega_{r}'(z)}\;,
\end{eqnarray}
from these equations, reconstructed the parameter $j(z)$.
\subsection{CPL model}\label{CPL}
Within this model, $E^{2}$ was determined replacing both $w(z)=w_{o}+w_{1}[z/(1+z)]$, where $w_{o}$, $w_{1}$ are real parameters 
and $Q(z)=0$ into Eqs. (\ref{Omegabs})-(\ref{EDFDEOmega})
\begin{eqnarray}\label{hubble_CPL}  
E^{2}&=&\Biggl[\Omega_{b,0}{(1+z)}^{3}+\Omega_{r,0}{(1+z)}^{4}+\Omega_{DM,0}{(1+z)}^{3} \nonumber\\
&& +\Omega_{DE,0}{(1+z)}^{3(1+w_{0}+w_{1})}{{\rm exp}}\biggl({\frac{-3w_{1}z}{1+z}}\biggr)\Biggr]\;.
\end{eqnarray}
Then, the following relation
\begin{equation}\label{w_cpl}
\omega'(z)=\frac{\omega_{1}}{(1+z)^{2}}\;,
\end{equation}
is substituted into Eqs. (\ref{q_values})-(\ref{q'_values}) and (\ref{jerk_definition}), from which reconstructed $q(z)$ and $j(z)$, respectively.\\
\subsection{XCPL model}\label{XCPL}
Here, firstly a coupled model was defined putting both $\omega=\omega_{o}+\omega_{1}(z/{1+z})$, where $\omega_{o}$, $\omega_{1}$ are real free
parameters and ${Q}(z)$ given by Eqs. (\ref{Interaction})-(\ref{CouplingIq}), into Eqs. (\ref{Omegabs})-(\ref{EDFDEOmega}).
The explicit form for $\Omega^{\star}_{DM}$ and $\Omega^{\star}_{DE}$ are reached, solving Eqs. (\ref{EDFDMOmega})-(\ref{EDFDEOmega}), 
respectively,
\begin{eqnarray}
\label{DM_XCPL}
\Omega^{\star}_{DM}(z)=(1+z)^{3}{\Omega_{DM,0}}{\rm exp}\biggl[{\frac{-z_{max}}{2}\sum_{n=0}^{2}\lambda_{n}I_{n}(z)}\biggr],\qquad\\
\label{DE_XCPL}
\Omega^{\star}_{DE}(z)=(1+z)^{3(1+\bar{w})}\biggl[{\Omega_{DE,0}}{\rm exp}\left(\frac{-3w_{1}z}{1+z}\right)+\;\quad\qquad\nonumber\\
\frac{z_{max}}{2}\Omega_{DM,0}{\rm exp}\left(\frac{3w_{1}}{1+z}\right)\sum_{n=0}^{2}\lambda_{n}S_{n}(z,\bar{w})\biggr]\;.\quad\qquad
\end{eqnarray}
The following average integrals have been defined
\begin{eqnarray}
\label{Average_In}\int_{0}^{z}\frac{T_{n}(\tilde{x})}{(1+\tilde{x})}d\tilde{x} & \approx & \frac{z_{max}}{2}I_{n}(z)\;,\\
\label{Average_Sn}\int_{0}^{z}\frac{T_{n}(\tilde{x})A(\tilde{x})B(\tilde{x})}{(1+\tilde{x})^{(1+3\bar{\omega})}}d\tilde{x} &\approx& 
\frac{z_{max}}{2}S_{n}(z,\bar{\omega})\;,
\end{eqnarray}
in where we have also defined the following expressions for all $n\in[0,2]$ (see Appendix \ref{integralIn} and \cite{Cueva-Nucamendi2012})
\begin{eqnarray}
A(\tilde{x})&=&{{\rm exp}}\left(\frac{-z_{max}}{2}{\sum_{n=0}^{2}}\lambda_{n}I_{n}(\tilde{x})\right)\;,\nonumber\\
\label{a1}\tilde{A}(\tilde{x})&=&{{\rm exp}}\left(\frac{-z_{max}}{2}{\sum_{n=0}^{2}}\lambda_{n}\tilde{I}_{n}(\tilde{x})\right)\;,\\
B(\tilde{x})&=&{{\rm exp}}\left(\frac{-3w_{1}}{1+\tilde{x}}\right),\quad\tilde{B}(\tilde{x})={{\rm exp}}\left(\frac{-3w_{1}}{a+b\tilde{x}}\right),\nonumber\\
I_{n}(z)&\equiv&\int_{-1}^{x} \frac{T_{n}(\tilde{x})}{(a+b\tilde{x})}d\tilde{x}\;,\nonumber\\
S_{n}(z,\bar{w})&\equiv&\int_{-1}^{x}\frac{T_{n}(\tilde{x})\tilde{A}(\tilde{x})\tilde{B}(\tilde{x})}{(a+b\tilde{x})^{(1+3\bar{w})}}d\tilde{x}\;,\nonumber
\end{eqnarray}
and the quantities,
\begin{eqnarray*}
x&\equiv& 2(z/z_{max})-1\;,\qquad \bar{\omega}\equiv \omega_{o}+\omega_{1}\;,\\
&&a\equiv1+\frac{z_{max}}{2}\;,\qquad  b\equiv\frac{z_{max}}{2}\;.
\end{eqnarray*}
where $z_{max}$ is the maximum value of $z$ in which the observations are possible so that $\tilde{x}\in [-1, 1]$ and 
$\vert T_{n}(\tilde{x})\vert\leq 1$.\\
Therefore, the function $E^{2}$ was constructed from Eqs. (\ref{Omegabs})-(\ref{Omegars}), (\ref{DM_XCPL})-(\ref{DE_XCPL}) and (\ref{hubble}). 
Similarly, Eq. (\ref{w_cpl}) is then substituted into Eqs. (\ref{q_values})-(\ref{q'_values}) and (\ref{jerk_definition}) to reconstruct $q(z)$ and $j(z)$, respectively.
\subsection{DR model}\label{DR}
Secondly a coupled model was modeled setting both $\omega(z)\equiv \omega_{o}+\omega_{1}z+\omega_{2}(2z^{2}-1)$, where $\omega_{o}$, $\omega_{1}$, 
$\omega_{2}$ are real parameters and ${Q}(z)$ given by Eqs. (\ref{Interaction})-(\ref{CouplingIq}), into Eqs. (\ref{EDFDMOmega})-(\ref{EDFDEOmega}).
The explicit form for $\Omega^{\star}_{DM}$ and $\Omega^{\star}_{DE}$ are reached, solving Eqs. (\ref{EDFDMOmega})-(\ref{EDFDEOmega}), 
respectively. In this way, $Q$ and $\omega$ were simultaneously reconstructed. For this model Eq. (\ref{DM_XCPL}) represents the solution of 
Eq. (\ref{EDFDMOmega}), and hence,  the solution of Eq. (\ref{EDFDEOmega}) was obtained using Eq. (\ref{a1}) and Appendix \ref{integralIn}, 
\begin{equation}\label{DE_DR}
\Omega^{\star}_{DE}(z)=C(z)+D(z)L(z)\int_{-1}^{x}{\sum_{n=0}^{2}\omega_{n}T_{n}(\tilde{x})}\frac{\tilde{A}(\tilde{x})F(\tilde{x})}{G(\tilde{x})}d\tilde{x}\;,
\end{equation}
where the following relations were defined
\begin{eqnarray*} 
C(z)&=&\Omega_{DE,0}(1+z)^{3}{\rm exp}\left(\frac{3z_{max}}{2}\sum_{n=0}^{2}\omega_{n}I_{n}(z)\right)\;,\\
D(z)&=&\frac{z_{max}}{2}\Omega_{DM,0}(1+z)^{3\left(\omega_{o}-\omega_{1}+\omega_{2}+1\right)}\;,\\
F(\tilde{x})&=&{\rm exp}\biggl[-3\biggl( \omega_{1}\left(1+\frac{z_{max}}{2}[1+\tilde{x}]\right)+\\
&& \omega_{2}{\left(1+\frac{z_{max}}{2}[1-\tilde{x}]\right)}^{2}\biggr)\biggr]\;,\\
G(\tilde{x})&=&\left[1+\frac{z_{max}}{2}(1+\tilde{x})\right]^{1-3(\omega_{1}-\omega_{2}-\omega_{o})}\;,\\
L(z)&=&{\rm exp}\biggl[3\left(\omega_{2}z^{2}+(\omega_{1}-2\omega_{2})z+\omega_{1}+\omega_{2}\right)\biggr]\;.
\end{eqnarray*}
Within this model, the function $E^{2}$ was constructed from Eqs. (\ref{Omegabs})-(\ref{Omegars}), (\ref{DM_XCPL}), (\ref{DE_DR}) and (\ref{hubble}). 
Furthermore, the relation
\begin{equation}\label{w_dr}
\omega'(z)=\omega_{1}+4z\omega_{2}\;,
\end{equation}
is replaced into Eqs. (\ref{q_values})-(\ref{q'_values}) and (\ref{jerk_definition}) to reconstruct $q(z)$ and $j(z)$, respectively. The basic analytical expre-ssions 
for $I_n(x)$ and $\tilde{I_{n}}(\tilde{x})$ are found in Appendix \ref{integralIn}.
\section{Current observational data and cosmological constraints.} \label{SectionAllTest}
In this section, we describe how we use the cosmological data currently available to test and constrain the parameter space of our 
models proposed.
\subsection{Type Ia supernovae data set.} \label{SNIa}
For the SNIa observations, we consider ``The Supernova Cosmology Project'' Union $2.1$ composed of $580$ SNIa data. 
The distance modulus $\mu(z,\mathbf{X})$, is defined as the difference between the apparent $(m)$ and absolute $(M)$ magnitudes, so that their 
observed and theoretical values are
\begin{eqnarray}\label{mus}
\mu^{{\rm obs}}&=&m^{{\rm obs}}-M \;,\\
{\mu}^{{\rm th}}(z,\mathbf{X})&\equiv& 5\log_{10}\left[\frac{D_L(z,\mathbf{X})}{{\rm Mpc}}\right]+\mu_{0}\;,
\end{eqnarray}
where $\mu_{0}=25-{\rm \log}_{10}H_{0}$, and the Hubble-free the luminosity distance $D_{L}$ \cite{Riess1998} in a flat cosmology is 
\begin{equation}\label{luminosity_distance1}
D_L(z,\mathbf{X})=(1+z) \int_{0}^{z}\frac{dz'}{H(z',\mathbf{X})}\;,
\end{equation}
in where $H(z,\mathbf{X})$ is the Hubble parameter, i.e., Eq. (\ref{hubble}) and, in general, $\mathbf{X}$ represents the model parameters
\begin{equation}\label{parametersm} 
\mathbf{X}\equiv(H_0,\Omega_{b,0},\Omega_{r,0},\Omega_{DM,0},\omega_{o},\omega_{1},\omega_{2},\lambda_{0},\lambda_{1},\lambda_{2}).
\end{equation}
The best fitting values of the parameters in a model are determined by the likelihood analysis of
\begin{equation}\label{ChiSN}
\chi^{2}_{\bf SN}(\mathbf{X},m_{0})\equiv\sum_{k = 1}^{580}\frac{\left[\mu^{obs}(z_{k})-\mu^{{\rm th}}(z_k,\mathbf{X})\right]^2}{{\sigma_k^2}(z_k)}\;,
\end{equation}
in where ${\sigma}(z_{k})$ is the corresponding $1\sigma$ error of distance modulus for each supernovae. The parameter $\mu_{0}$ is a nuisance 
parameter. According to \cite{nesseris1}, $\chi^{2}$ is expanded as 
\begin{equation}\label{ChiSN1}
{{\chi}}^{2}_{\bf SN}(\mathbf{X})=A_{1}-2\mu_{0}B_{1}+{\mu_{0}}^{2}C_{1}\;,
\end{equation}
where
\begin{eqnarray}\label{ChiSN2}
A_{1}&=&\sum_{k=1}^{580}\frac{\left[\mu^{obs}(z_k)-\mu^{th}(z_k,\mathbf{X})\right]^2}{{\sigma_k}^2(z_k)}\;,\nonumber\\
B_{1}&=&\sum_{k=1}^{580}\frac{\left[\mu^{obs}(z_k)-\mu^{th}(z_k,\mathbf{X})\right]}{{\sigma_k}^2(z_k)}\;,\nonumber\\
C_{1}&=&\sum_{k=1}^{580}\frac{1}{{\sigma_{k}}^2(z_k)}\;.
\end{eqnarray}
The Eq.(\ref{ChiSN1}) has a minimum for $\mu_{0}=B_{1}/C_{1}$ at
\begin{equation}\label{ChiSNIa}
{\tilde{\chi}}^{2}_{\bf SN}(\mathbf{X})=A_{1}(\mathbf{X})-\frac{B_{1}^{2}(\mathbf{X})}{C_{1}(\mathbf{X})}\;\;.
\end{equation}
Since ${{\chi}}^{2}_{SN,min}={\tilde{\chi}}^{2}_{SN,min}$, instead minimizing ${{\chi}}^{2}_{SN}$ we will minimize ${\tilde{\chi}}^{2}_{SN}$,
which is independent of $\mu_{0}$.
\subsection{BAO data sets} \label{BAO}
Eisenstein et al.\cite{Eisenstein1998}, first found a well-detected peak of the imprint of the recombination-epoch acoustic oscillations 
in the large-scale correlation function at $100h^{-1}$Mpc ($h\equiv H_{0}/100Kms^{-1}Mpc^{-1}$) separation measured from a spectroscopic 
sample of $46,748$ luminous red galaxies of the SDSS. Also, Percival et al.\cite{ReidSDSS}, investigated the clustering of 
galaxies within the spectroscopic SDSS-DR$7$ galaxy sample including, the luminous red galaxy,
main samples, and also the $2$-degree Field Galaxy Redshift Survey ($2$dFGRS) data (in total $893,319$ galaxies) observed BAO 
in power spectrum of matter fluctuations after the epoch of recombination on large scales. This allowed to detect the BAO signal at $z=0.2$ and $z=0.35$.\\
Eisenstein first and Percival after constructed an effective distance ratio $D_{v}(z)$, which encodes the visual distortion of a 
spherical object due to the non-Euclidianity of a FRW spacetime, defined as
\begin{eqnarray}\label{Dv}
D_{v}(z,\mathbf{X})&\equiv&\frac{1}{H_{0}}\left[(1+z)^{2}{D_{A}}^{2}(z)\frac{cz}{E(z)}\right]^{1/3},\nonumber\\
&=&\frac{c}{H_{0}}\left[\left(\int_{0}^{z}\frac{dz'}{E(z',\mathbf{X})}\right)^{2}\frac{z}{E(z,\mathbf{X})}\right]^{1/3}
\end{eqnarray}
where $D_{A}(z)$ is the proper (not comoving) angular diame-ter distance, which has the following definition
\begin{eqnarray} \label{DA}
D_{A}(z,\mathbf{X}) &\equiv& \frac{c}{(1+z)}{\int}^{z}_{0}\frac{dz'}{H(z',\mathbf{X})}\;.
\end{eqnarray}
The comoving sound horizon size is defined by
\begin{equation} \label{hsd} 
r_{s}(a)\equiv c\int^{a}_{0}\frac{c_{s}(a')da'}{{a'}^{2}H(a')}\;\;,
\end{equation}
being $c_{s}(a)$ the sound speed of the photon-baryon fluid
\begin{equation} \label{vsd}
c_{s}^{2}(a) \equiv \frac{\delta P}{\delta \rho}=\frac{1}{3}\left[\frac{1}{1+(3\Omega_{b}/4\Omega_{r})a}\right]\;\;.
\end{equation}
Considering the Eqs. (\ref{hsd}) and (\ref{vsd}) for a $z$ given   
\begin{equation}\label{rs}
r_{s}(z)=\frac{c}{\sqrt3}{\int}^{1/(1+z)}_{0}\frac{da}{a^{2}H(a)\sqrt{1+(3\Omega_{b,0}/4\Omega_{\gamma,0})a}}\;\;\;,
\end{equation}
where $\Omega_{b,0}$ and $\Omega_{\gamma,0}$ are the present-day baryon and photon density parameters, respectively. In this paper, we have fixed 
$\Omega_{\gamma,0}=2.469\times10^{-5}h^{-2}$, $\Omega_{b,0}=0.02246h^{-2}$, given by the WMAP$7$ data \cite{Komatsu2011} and 
$\Omega_{r,0}=\Omega_{\gamma,0}(1+0.2271N_{eff})$ with $N_{eff}$ the effective number of neutrino species (here the standard value, $N_{eff}=3.04$ 
was chosen \cite{Komatsu2011}). The peak position of the BAO depends on the ratio of $D_{v}(z)$ to the sound horizont size at the drag epoch 
(where baryons were released from photons) $z_{d}$, which was obtained by using a fitting formula \cite{Eisenstein1998}:
\begin{equation}\label{zd}
z_{d}=\frac{1291(\Omega_{M,0}h^{2})^{0.251}}{1+0.659(\Omega_{M,0}h^{2})^{0.828}}\left(1+b_{1}(\Omega_{b,0}h^{2})^{b_{2}}\right)\;,
\end{equation}
where $\Omega_{M,0}=\Omega_{DM,0} + \Omega_{b,0}$ and
\begin{eqnarray*} 
b_{1}&=&0.313(\Omega_{M,0}h^{2})^{-0.419}\left[1+ 0.607(\Omega_{M,0}h^{2})^{0.674}\right]\;,\\
b_{2}&=&0.238(\Omega_{M,0}h^{2})^{0.223}\;.
\end{eqnarray*}
The distance radio $d_{z}$ at $z=0.2$ and $z=0.35$ extracted from the SDSS and $2$dFGRS \cite{Eisenstein1998,SDSS,ReidSDSS}, are listed in
Table \ref{tableBAO},
\begin{equation} \label{dvalues}
d_{0.20}(\mathbf{X})=\frac{r_{s}(z_{d})}{D_{V}(0.2,\mathbf{X})}\;,\hspace{0.3cm} d_{0.35}(\mathbf{X})=\frac{r_{s}(z_{d})}{D_{V}(0.35,\mathbf{X})}\;,
\end{equation}
where $r_{s}(z_{d}, \mathbf{X})$ is the comoving sound horizont size at the baryon drag epoch.
Using the data of Table \ref{tableBAO}, then the inverse covariance matrix of BAO \cite{ReidSDSS} is
\begin{table}
\centering
\begin{tabular}{| c | c |}
 \hline
 $z$ & $d_{z}$  \\
 \hline
 $0.20$ & $0.1905\pm0.0061$ \\
 \hline
 $0.35$ & $0.1097\pm0.0036$ \\
 \hline
\end{tabular}
\caption{Summary of SDSS-$2$dFGS BAO data set \cite{ReidSDSS}.}
 \label{tableBAO}
\end{table}
\begin{equation}\label{MatrixBAO}
C^{-1}_{\bf SDSS}=\left(\begin{array}{cc}
 +30124  -17227 \\
 -17227  +86977 \\
\end{array} \right)\;.
\end{equation}
The errors of the SDSS-$2$dFGS BAO data are contained in $C^{-1}_{\bf SDSS}$. The $\chi^{2}$ of the SDSS-$2$dFGS BAO data set is 
\begin{equation}\label{X2BAO1}
\chi_{\bf SDSS}^{2}(\mathbf{X})=\left({\Delta d}_{i}\right)\left(C^{-1}_{\bf SDSS}\right)_{ij}\left({\Delta d}_{j}\right)^{t},
\end{equation}
where ${\Delta d}_{i}=d^{th}_{i}(\mathbf{X})-d^{obs}_{i}$ is a column vector 
\begin{equation}
d^{th}_{i}(\mathbf{X})-d^{obs}_{i}=\left(\begin{array}{rl}                               
 d_{0.20}(\mathbf{X})-0.1905 \\
 d_{0.35}(\mathbf{X})-0.1097 \\
\end{array}\right)\;,
\end{equation}
and ``t'' denotes its transpose.\\
On the other hand, in the low redshift region $(z<1)$, Beutler et al. studied the large-scale correlation function of the $6$dFGS and detected a BAO signal 
\cite{Beutler2011}. This detection allowed to constrain the distance-redshift relation $D_{v}(z)$
at $z_{eff}=0.106$, $D_{v}(z_{eff})=456\pm27$ Mpc and a measurement of the distance ratio, $d_{z_{eff}}={r_{s}(z_{d})}/{D_{v}(z_{eff})}=0.336\pm0.015$, where $r_{s}(z_{d})$ is the sound horizon at the drag epoch $z_{d}$.\\
The $\chi^{2}$ function of the 6dFGS BAO data set is given by
\begin{equation}\label{X2BAO2}
\chi_{\bf 6dFGS}^{2}(\mathbf{X})=\left(\frac{d_{z}-0.336}{0.015}\right)^{2}\;\;.
\end{equation}
Also, in \cite{Blake2011}, Blake et al. presented new measurements of the BAO at $z=0.44, 0.6$, and $0.73$ in the galaxy correlation 
function of the final data set of the WiggleZ. This sample included a total of $N=158,741$ galaxies in the range $0.2<z<1.0$. Here, the acoustic parameter $A$, 
was defined as \cite{Eisenstein1998}:
\begin{eqnarray}\label{A}
A(z)&\equiv& D_{v}(z)\sqrt{\Omega_{M,0}}\left(\frac{H_{0}}{cz}\right)\;,\nonumber\\
&=&\sqrt{\Omega_{M,0}}\left(\left(\int_{0}^{z}\frac{dz'}{E(z',\mathbf{X})}\right)^{2}\frac{1}{z^{2}E(z,\mathbf{X})}\right)^{1/3}
\end{eqnarray}
Using the BAO data showed into Table \ref{tableBAO_WiggleZ}, the inverse covariance matrix of WiggleZ BAO \cite{Blake2011} is given by
 \begin{table}
\centering
\begin{tabular}{| l | c |}
\hline
 $z_{eff}$ & $A(z)$  \\
 \hline
 $0.44$ & $0.474\pm0.034$\\
 \hline
 $0.60$ & $0.442\pm0.020$\\
 \hline 
 $0.73$ & $0.424\pm0.021$\\
\hline
\end{tabular}
\caption{Summary of the effective redshifts $z_{eff}$ and the corresponding measurements for the acoustic parameter $A(z)$ of the 
WiggleZ BAO data \cite{Blake2011}.}
\label{tableBAO_WiggleZ}
\end{table}  
\begin{equation} \label{MatriWiggleZBAO}
C^{-1}_{\bf WiggleZ}=\left(\begin{array}{ccc}           
+1040.3  -807.5  +336.8\\
-807.5  +3720.3  -1551.9\\
+336.8  -1551.9  +2914.9\\
\end{array} \right)\;.
\end{equation}
Then, define the $\chi^{2}$ function of the WiggleZ BAO data  
\begin{equation}\label{X2BAO3}
\chi_{\bf WiggleZ}^{2}(\mathbf{X})=\left({\Delta A}_{i}\right)\left(C^{-1}_{\bf WiggleZ}\right)_{ij}\left({\Delta A}_{j}\right)^{t},
\end{equation}
where ${\Delta A}_{i}=A^{th}_{i}(\mathbf{X})-A^{obs}_{i}$ is a column vector 
\begin{equation}
A^{th}_{i}(\mathbf{X})-A^{obs}_{i}=\left(\begin{array}{rl}                               
 A(0.44,\mathbf{X})-0.474 \\
 A(0.60,\mathbf{X})-0.442 \\
 A(0.73,\mathbf{X})-0.424 \\
\end{array}\right)\;,
\end{equation}
and ``t'' denotes its transpose.\\
Since Eqs. (\ref{X2BAO1}), (\ref{X2BAO2}) and (\ref{X2BAO3}), the total $\chi^{2}$ for the BAO data sets is constructed
\begin{equation}
\chi_{\bf BAO}^{2}=\chi_{\bf SDSS}^{2}+\chi_{\bf 6dFGS}^{2}+\chi_{\bf WiggleZ}^{2}\;\;.
\end{equation}
\subsection{CMB data set} \label{CMB}
The Union $2.1$ SNIa and BAO data sets contain information about the universe at low redshifts, we now include WMAP $7$ data \cite{Komatsu2011} to probe the 
entire expansion history up to the last scattering surface.
The shift parameter ${\rm {\bf R}}$ is provided by \cite{Bond-Tegmark1997}
\begin{eqnarray}\label{Shiftparameter}
{\rm {\bf R}}(z_{*},\mathbf{X})\equiv \frac{H_{0}}{c}\sqrt{\Omega_{M,0}}(1+z_{*})D_{A}(z_{*},\mathbf{X}),\nonumber\\
=\sqrt{\Omega_{M,0}}{\int}^{z_{*}}_{0}\frac{d\tilde{y}}{E(\tilde{y})}\,,\hspace{2cm}
\end{eqnarray}
where the distance $D_{A}$ and $E(\tilde{y})$ are given by Eqs. (\ref{DA}) and (\ref{hubble}), respectively. 
Then, the redshift $z_{*}$ (the decoupling epoch of photons) was obtained using the following fitting function \cite{Hu-Sugiyama1996}
\begin{equation}\label{Redshift_decoupling}
z_{*}=1048\biggl[1+0.00124(\Omega_{b,0}h^{2})^{-0.738}\biggr]\biggl[1+g_{1}(\Omega_{M,0}h^{2})^{g_{2}}\biggr]\;,\;\;
\end{equation}
where $\Omega_{M,0}=\Omega_{DM,0}+\Omega_{b,0}$, moreover, $g_{1}$ and $g_{2}$ are 
\begin{equation}\label{g1g2}
g_{1}=\frac{0.0783(\Omega_{b,0}h^{2})^{-0.238}}{1+39.5(\Omega_{b,0}h^{2})^{0.763}}\;,\hspace{0.3cm}g_{2}=\frac{0.560}{1+21.1(\Omega_{b,0}h^{2})^{1.81}}.
\end{equation}
An angular scale $l_{A}$ for the sound horizon at decoupling epoch was defined as
\begin{equation}\label{Acoustic_scale}
l_{A}(\mathbf{X})\equiv(1+z_{*})\frac{\pi D_{A}(z_{*},\mathbf{X})}{r_{s}(z_{*},\mathbf{X})}\,,\hspace{1cm}
\end{equation}
where $r_{s}(z_{*},\mathbf{X})$ is the comoving sound horizon at $z_{*}$, and is given by Eq. (\ref{rs}).
The maximum likelihood values according WMAP $7$ data \cite{Komatsu2011} are given in Table \ref{tableCMB}.\\  
\begin{table}
\centering
\begin{tabular}{| c | c |}
 \hline
 $l_{A}(z_{*})$ & $302.09 \pm 0.76$ \\
 \hline
 $R(z_{*})$ & $1.725 \pm 0.018$ \\
 \hline
 $z_{*}$ & $1091.3 \pm 0.91$ \\
 \hline
\end{tabular}
\caption{The distance priors given by \cite{Komatsu2011}.}
\label{tableCMB}
\end{table}
Then, following \cite{Komatsu2011} the $\chi^{2}$ for the CMB data is
\begin{equation}\label{X2CMB}
\chi_{\bf CMB}^{2}(\mathbf{X})=\left({\Delta x}_{i}\right)\left(C^{-1}_{\bf CMB}\right)_{ij}\left({\Delta x}_{j}\right)^{t},
\end{equation}
where ${\Delta x}_{i}=x^{th}_{i}(\mathbf{X})-x^{obs}_{i}$ is a column vector 
\begin{equation}
x^{th}_{i}(\mathbf{X})-x^{obs}_{i}=\left(\begin{array}{cc}
 l_{A}(z_{*})-302.09 \\
 R(z_{*})-1.725   \\
 \;\;z_{*}-1091.3 \\
\end{array}\right)\;,
\end{equation}
``t'' denotes its transpose and $(C^{-1}_{\bf CMB})_{ij}$ is the inverse covariance matrix (\cite{Komatsu2011})
given by
\begin{equation}\label{MatrixCMB}
C^{-1}_{\bf CMB}\equiv\left(
\begin{array}{ccc}
+2.3050&+29.6980&-1.3330\\
+29.698&+6825.27&-113.18\\
-1.3330&-113.180&+3.4140\\
\end{array}\right)\;.
\end{equation}
The errors for the CMB data are contained in $C^{-1}_{\bf CMB}$.
\begin{table*}[!htb]
\begin{tabular}{>{\centering\arraybackslash}m{7.8cm} >{\arraybackslash}m{10cm}}
{\begin{tabular}{| c  | c  | c  |}
\hline
 $z$   & $H(z)$ &  $1\sigma$\\
\hline
$0.0$  & $74.2$&  $\pm3.6$\\
$0.10$ & $69$  &  $\pm12$\\
$0.17$ & $83$  &  $\pm8$\\
$0.27$ & $77$  &  $\pm14$\\
$0.40$ & $95$  &  $\pm17$\\
$0.48$ & $97$  &  $\pm60$\\
$0.88$ & $90$  &  $\pm40$\\
$0.90$ & $117$ &  $\pm23$\\
$1.30$ & $168$ &  $\pm17$\\
$1.43$ & $177$ &  $\pm18$\\
$1.53$ & $140$ &  $\pm14$\\
$1.73$ & $202$ &  $\pm40$\\ 
\hline\end{tabular}\caption{Shows the observational $H(z)$ data \cite{Hubble2009-Stern2010}.}\label{tableOHD}} & 
{\begin{tabular}{|c|@{\extracolsep{0mm}\ }c@{ }|}
\hline
Parameters&Constant Priors\\[0.2mm]
\hline
$\Omega_{DM,0}$&$[0,0.7]$\\[0.2mm]
$H_{0}(kms^{-1}{Mpc}^{-1})$&$[20,140]$\\[0.2mm]
${\lambda}_{0}$&$[-1.5\times10^{+6},+1.5\times10^{+6}]$\\[0.2mm]
${\lambda}_{1}$&$[-1.5\times10^{+6},+1.5\times10^{+6}]$\\[0.2mm]
${\lambda}_{2}$&$[-1.5\times10^{+6},+1.5\times10^{+6}]$\\[0.2mm]
$\omega_{o}$&$[-2.0,-0.3]$\\[0.2mm]
$\omega_{1}$&$[-1.0,+1.0]$\\[0.2mm]
$\omega_{2}$&$[-10,+10]$\\[0.2mm]
\hline
\end{tabular} 
\caption{Shows the priors on the parameter space.\hspace{2.7cm}}\label{Priors}}
\end{tabular}
\end{table*}  
\subsection{Observational Hubble data (H)}\label{OHD}
In \cite{Jimenez2002, Jimenez2003, Simon2005}, the authors established that it is possible to compute the observational Hubble data, by using di-fferential ages of
galaxies through the measuring of $dz/dt$. Then, the Hubble parameter was expressed in terms of the differential ages as 
\begin{equation}\label{Hubble}
{\bf H}(z)=-\frac{1}{1+z}\frac{dz}{dt}\;. 
\end{equation}
Simon et al.\cite{Simon2005} found Hubble data over the redshift range $[0, 1.8]$. In \cite{Hubble2009-Stern2010}, the authors found new data of the Hubble 
parameter at $z \in [0.35,1]$ from SPICES and VVDS galaxy surveys, respectively, which are listed in Table \ref{tableOHD}. 
In addition, in \cite{Gaztanaga}, the authors took the BAO scale as a standard ruler in the radial direction and found three more additional data:
$H(z=0.24)$ $=79.69\pm2.32$, $H(z=0.34)=83.80\pm2.96$ and $H(z=0.43)=86.45\pm3.27$ (in units of $Kms^{-1}Mpc^{-1}$).\\
The $\chi^2$ for the observational Hubble data is \cite{Lazkoz2007}
\begin{equation}\label{X2OHD}
\chi^2_{H}(\mathbf{X})\equiv\sum_{i = 1}^{15}\frac{\left[H^{{\rm th}}(\mathbf{X},z_{i},)-H^{obs}(z_{i})\right]^2}{\sigma^2(z_{i})}\;\;,
\end{equation}
where $\mathbf{X}$ represents the parameters of the model, $H^{{\rm th}}$ is the theoretical value for the Hubble parameter, 
$H^{obs}$ is the observed value, and $\sigma(z_{i})$ is the standard deviation measurement uncertainty. Here the summation is over the 15 observational Hubble data  
at $z_{i}$. This test has been already used to constrain several models in \cite{Lu2009, Xu2010, Feng2011}.\\
Therefore, to fit our models with observations, we use all the data sets described above.\\
The best fitted parameters are obtained by minimizing 
\begin{equation}\label{TotalChi}
{{{\rm {\bf\tilde{\chi}}^{2}}}}={{{\rm {\bf\tilde{\chi}}^{2}}}}_{\bf SN}+{{{\rm {\bf {\chi}}^{2}}}}_{\bf BAO}+{{{\rm {\bf {\chi}}^{2}}}}_{\bf CMB}+
{{{\rm {\bf {\chi}}^{2}}}}_{\bf H}\;\;.
\end{equation}
From (\ref{TotalChi}), we will construct the total probability density function ${{\rm {\bf pdf}}}$ as 
\begin{equation}\label{TotalexpChi}
{{\rm {\bf pdf}}}(\mathbf{X})=\rm{A}{{\rm e}}^{-{\tilde{\chi}}^{2}/2}\;\;.
\end{equation}
where $\rm A$ is a integration constant.
\subsection{Constant Priors}\label{Values}
In this work, we have assumed that baryonic matter ($b$) and radiation ($r$) are not coupled to $DE$ or $DM$ and are separately 
conserved \cite{Koyama2009-Brax2010}. In this regard, we believe that the intensity of the interaction, ${\rm I}_{\rm Q}$, is not affected by the values of 
$\Omega_{b,0}$ and $\Omega_{r,0}$, respectively. Due to it, in this paper, we have fixed: $\Omega_{\gamma,0}=2.469\times10^{-5}h^{-2}$ and 
$\Omega_{b,0}=0.02246h^{-2}$, given by WMAP $7$ data \cite{Komatsu2011}. Then, based in these assumptions, construct ${{\rm {\bf pdf}}}$ for each of our models. 
The priors on the parameters space are given in Table \ref{Priors}, which were used in all the observational tests of our models.
They have allowed us to compute the best fitting values of the free parameters.
\begin{figure*}[!htb]
 \includegraphics[width=15.9cm,height=7.2cm]{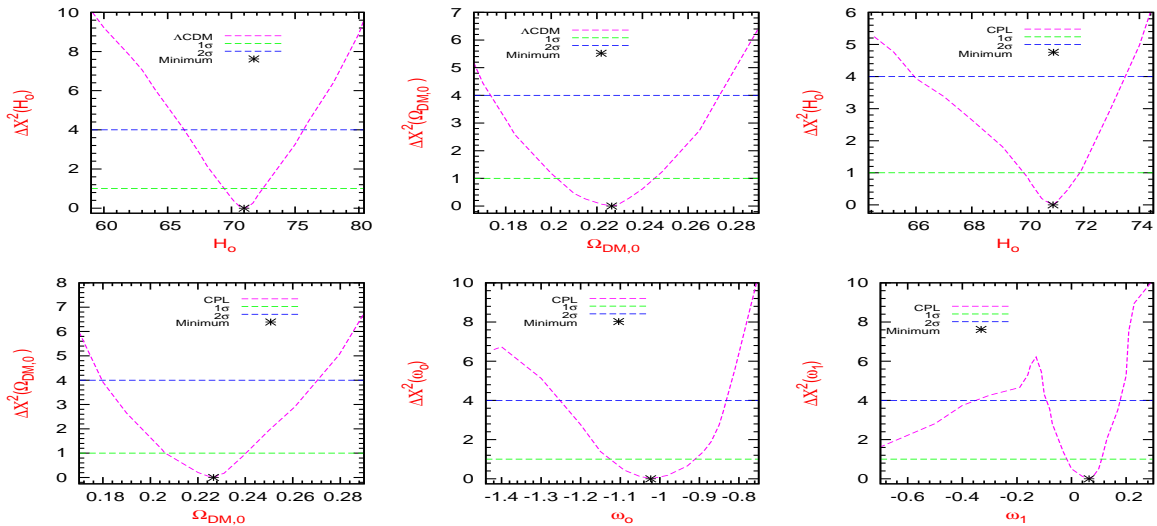}\\
 \caption{(color online) Shows the one-dimension probability contours for each of the parameters of our noncoupled models.\\
 In each panel the black star denotes the best fitting value of the parameters. The $1\sigma$ and $2\sigma$ represent the errors.}\label{LCDM-CPL_CI}
\end{figure*}
\begin{figure*}[!htb]
 \includegraphics[width=8.4cm,height=7.6cm]{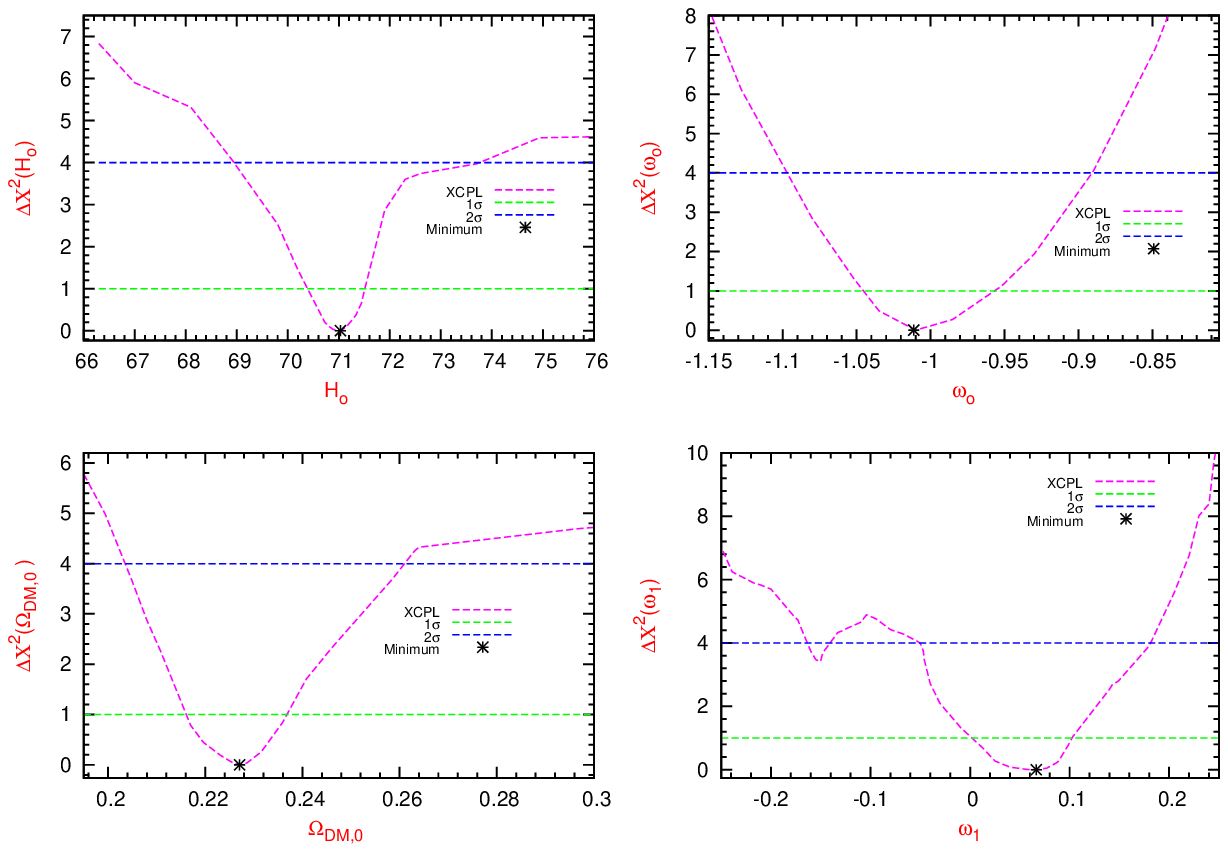}\includegraphics[width=8.4cm,height=7.6cm]{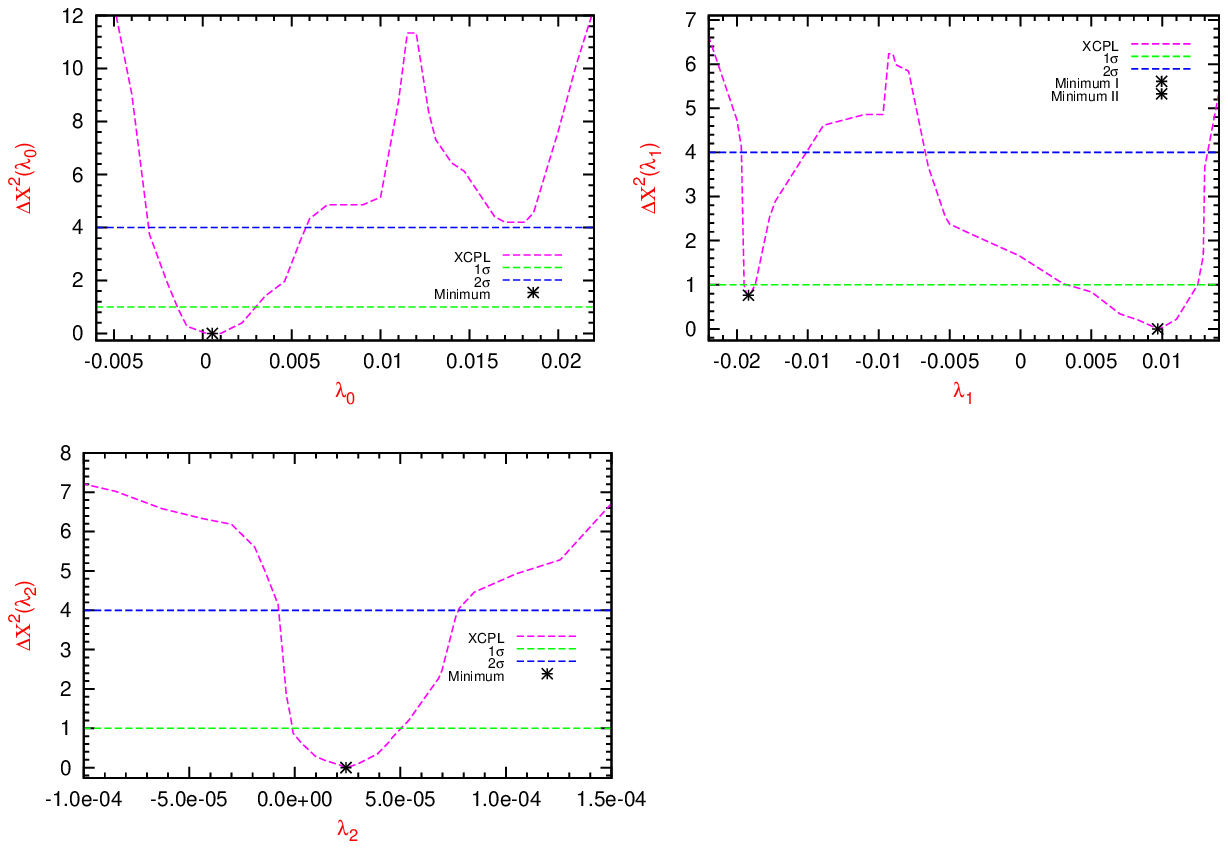}\\
 \caption{(color online) Displays the one-dimension probability contours for the parameters of the XCPL model and their constraints at $1\sigma$ 
 and $2\sigma$, respectively. In each panel the black star denotes the best fitting parameter.}\label{XCPL_CI}
\end{figure*} 
\begin{figure*}[!htb]
 \includegraphics[width=8.4cm,height=7.6cm]{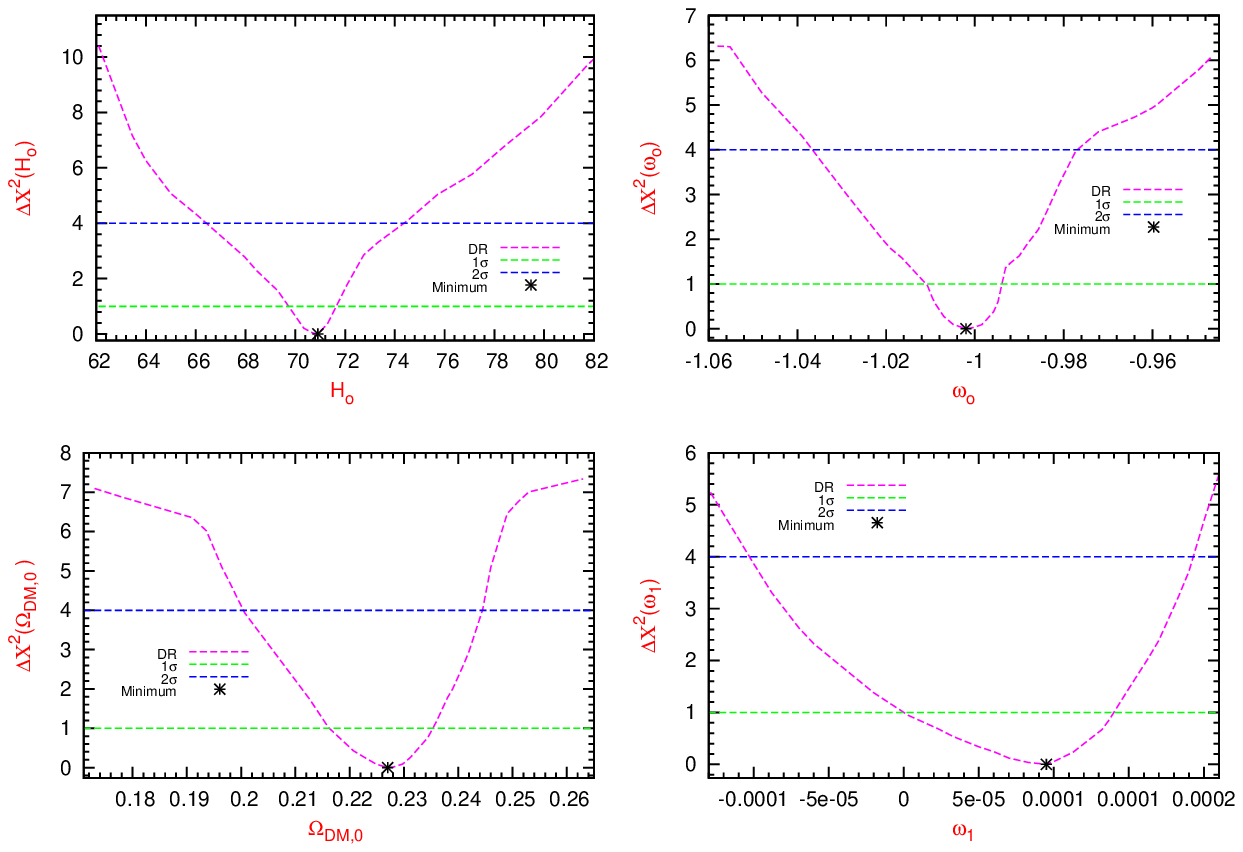}\includegraphics[width=8.4cm,height=7.6cm]{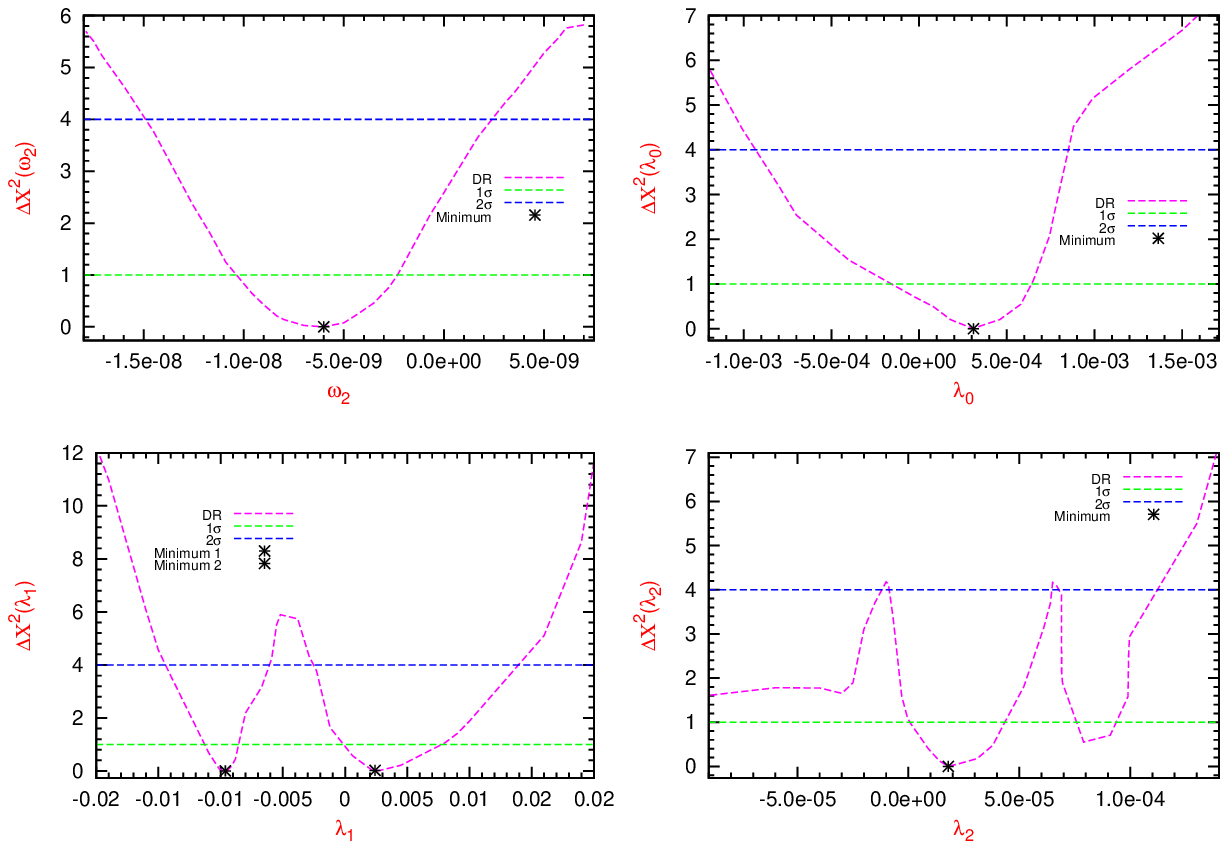}\\
 \caption{(color online) Displays the one-dimension probability contours for the parameters of the DR model and their constraints at $1\sigma$ 
 and $2\sigma$, respectively. In each panel the black star denotes the best fitting parameter.}\label{DR_CI}
\end{figure*} 
\begingroup
\squeezetable
\begin{table*}[!hbtp]
\centering
\caption{Shows the best fitting cosmological parameters for each model, and their constraints at $1\sigma$ and $2\sigma$, obtained from an analysis of 
Union $2.1$ SNIa+BAO+CMB+H data sets.}\label{Bestfits}
\begin{tabular}{| c | c | c | c |}
\hline\hline
 Parameters&$\Lambda$CDM&CPL&XCPL(I)\\
 \hline\hline
${\lambda}_{0}$&$N/A$&$N/A$&${+0.536\times10^{-3}}^{+2.356\times10^{-3}+5.249\times10^{-3}}_{-2.036\times10^{-3}-3.643\times10^{-3}}$\\[0.4mm]
${\lambda}_{1}$&$N/A$&$N/A$&${+9.68\times10^{-3}}^{+2.720\times10^{-3}+3.420\times10^{-3}}_{-6.470\times10^{-3}-16.376\times10^{-3}}$\\[0.4mm]
${\lambda}_{2}$&$N/A$&$N/A$&${+2.43\times10^{-5}}^{+2.641\times10^{-5}+5.284\times10^{-5}}_{-2.644\times10^{-5}-3.287\times10^{-5}}$\\[0.4mm]
$\omega_{o}$&$-1.0$&${-1.0231}^{+0.1118+0.1916}_{-0.1033-0.2299}$&${-1.0040}^{+0.0453+0.1140}_{-0.0435-0.0947}$\\[0.4mm]
$\omega_{1}$&$N/A$&${+0.0642}^{+0.0465+0.1129}_{-0.0842-0.1592}$&${+0.0665}^{+0.0353+0.1148}_{-0.0638-0.1183}$\\[0.4mm]
$\omega_{2}$&$N/A$&$N/A$&$N/A$\\[0.4mm]
$\Omega_{DM,0}$&${+0.2265}^{+0.0187+0.0475}_{-0.0241-0.0533}$&${+0.2266}^{+0.0194+0.0440}_{-0.0205-0.0470}$&${+0.2271}^{+0.0094+0.0314}_{-0.0132-0.0227}$\\[0.4mm]
$H_{0}(kms^{-1}{Mpc}^{-1})$&${+70.9963}^{+1.5049+4.6282}_{-1.6108-4.6937}$&${+70.9060}^{+0.9572+2.5967}_{-1.0491-4.9753}$
&${+71.0321}^{+0.4785+2.7324}_{-0.6659-2.1272}$\\[0.4mm]
\hline\hline$\tilde{\chi}^{2}_{min}$&$567.5023$&$559.68457$&$557.87849$\\[0.4mm]
\hline
\end{tabular}
\centering\begin{tabular}{| c | c | c | c |}
\hline\hline
 Parameters&XCPL(II)&DR(1)&DR(2)\\
 \hline\hline
${\lambda}_{0}$&${+0.536\times10^{-3}}^{+2.356\times10^{-3}+5.249\times10^{-3}}_{-2.036\times10^{-3}-3.643\times10^{-3}}$
&${+3.1\times10^{-4}}^{+3.3243\times10^{-4}+5.4365\times10^{-4}}_{-4.7626\times10^{-4}-12.4271\times10^{-4}}$&
${+3.1\times10^{-4}}^{+3.3243\times10^{-4}+5.4365\times10^{-4}}_{-4.7626\times10^{-4}-12.4271\times10^{-4}}$\\[0.4mm]
${\lambda}_{1}$&${-19.2\times10^{-3}}^{+0.5\times10^{-3}+4.0\times10^{-3}}_{-0.4\times10^{-3}-0.5\times10^{-3}}$
&${-9.62\times10^{-3}}^{+1.07\times10^{-3}+3.5825\times10^{-3}}_{-1.5925\times10^{-3}-4.78\times10^{-3}}$&
${+2.4\times10^{-3}}^{+5.314\times10^{-3}+11.4036\times10^{-3}}_{-2.9339\times10^{-3}-4.9268\times10^{-3}}$\\[0.4mm]
${\lambda}_{2}$&${+2.43\times10^{-5}}^{+2.641\times10^{-5}+5.284\times10^{-5}}_{-2.644\times10^{-5}-3.287\times10^{-5}}$
&${+1.8\times10^{-5}}^{+2.5496\times10^{-5}+4.6683\times10^{-5}}_{-1.7264\times10^{-5}-2.6510\times10^{-5}}$&
${+1.8\times10^{-5}}^{+2.5496\times10^{-5}+4.6683\times10^{-5}}_{-1.7264\times10^{-5}-2.6510\times10^{-5}}$\\[0.4mm]
$w_{0}$&${-1.0040}^{+0.0453+0.1140}_{-0.0435-0.0947}$&${-1.0020}^{+0.0082+0.0252}_{-0.0092-0.0346}$&${-1.0020}^{+0.0082+0.0252}_{-0.0092-0.0346}$\\[0.4mm]
$w_{1}$&${+0.0665}^{+0.0353+0.1148}_{-0.0638-0.1183}$&${+9.5\times10^{-5}}^{+4.4857\times10^{-5}+9.7620\times10^{-5}}_{-9.7914\times10^{-5}-19.801\times10^{-5}}$&
${+9.5\times10^{-5}}^{+4.4857\times10^{-5}+9.7620\times10^{-5}}_{-9.7914\times10^{-5}-19.801\times10^{-5}}$\\[0.4mm]
$w_{2}$&$N/A$&${-6.0\times10^{-9}}^{+3.6395\times10^{-9}+8.4502\times10^{-9}}_{-4.4308\times10^{-9}-8.9306\times10^{-9}}$&
${-6.0\times10^{-9}}^{+3.6395\times10^{-9}+8.4502\times10^{-9}}_{-4.4308\times10^{-9}-8.9306\times10^{-9}}$\\[0.4mm]
$\Omega_{DM,0}$&${+0.2271}^{+0.0094+0.0314}_{-0.0132-0.0227}$&${+0.2270}^{+0.0085+0.0176}_{-0.0114-0.0269}$&${+0.2270}^{+0.0085+0.0176}_{-0.0114-0.0269}$\\[0.4mm]
$H_{0}(kms^{-1}{Mpc}^{-1})$&${+71.0321}^{+0.4785+2.7324}_{-0.6659-2.1272}$&${+70.91}^{+0.7514+3.4634}_{-1.1901-4.5339}$
&${+70.91}^{+0.7514+3.4634}_{-1.1901-4.5339}$\\[0.4mm]
\hline\hline$\tilde{\chi}^{2}_{min}$&$558.04200$&$556.85491$&$556.27088$\\[0.4mm]
\hline\hline
\end{tabular}
\end{table*}
\endgroup
\begingroup
\squeezetable
\begin{table*}[!htbp]
\centering
\caption{Shows the best fitting parameters today for each model, $q_{o}$, $j_{o}$, $I_{o}$, $\omega_{o}$ and $\Omega_{DM,0}$, and their errors at $1\sigma$ and 
$2\sigma$, obtained from a combination of data.}\label{cosmological_state}
\begin{ruledtabular}
\begin{tabular}{| c | c | c | c | c | c |}
Models&$q_{o}$&$j_{o}$&$I_{o}\times10^{4}$&$\omega_{o}$&$\Omega_{DM,0}$\\
\hline\hline
$\Lambda$CDM&${-0.5932}^{+0.0253+0.0633}_{-0.0331-0.0700}$&${-1.00017}^{+1\times10^{-5}+2\times10^{-5}}_{-0.0-2\times10^{-4}}$&$0.0$&$-1.0$
&${-0.2265}^{+0.0187+0.0475}_{-0.0241-0.0533}$\\[0.35mm]
CPL&${-0.6181}^{+0.1396+0.2604}_{-0.1454-0.3265}$&${-1.1478}^{+0.2891+0.3985}_{-0.3092-0.8395}$&$0.0$&${-1.0231}^{+0.1118+0.1916}_{-0.1033-0.2299}$
&${+0.2266}^{+0.0140+0.0440}_{-0.0205-0.0470}$\\[0.35mm]
XCPL(I)&${-0.5967}^{+0.0622+0.1622}_{-0.0668-0.1362}$&${-1.0892}^{+0.1041+0.2047}_{-0.0813-0.2169}$
&${-5.1141}^{+23.299+51.964}_{-20.093-36.098}$&${-1.0040}^{+0.0453+0.1140}_{-0.0435-0.0947}$&${+0.2271}^{+0.0094+0.0314}_{-0.0132-0.0227}$\\[0.35mm]
XCPL(II)&${-0.5967}^{+0.0621+0.1622}_{-0.0669-0.1363}$&${-1.0793}^{+0.1047+0.2045}_{-0.0835-0.2224}$
&${+5.1141}^{+23.299+51.964}_{-20.093-36.098}$&${-1.0040}^{+0.0453+0.1140}_{-0.0435-0.0947}$&${+0.2271}^{+0.0094+0.0314}_{-0.0132-0.0227}$\\[0.35mm]
DR(1)&${-0.5945}^{+0.0202+0.0474}_{-0.0250-0.0698}$&${-1.0036}^{+0.0262+0.0783}_{-0.0306-0.1203}$
&${+2.9200}^{+3.0693+4.9697}_{-4.5900-12.162}$&${-1.0020}^{+0.0082+0.0252}_{-0.0092-0.0346}$&${+0.2270}^{+0.0085+0.0176}_{-0.0114-0.0269}$\\[0.35mm]
DR(2)&${-0.5945}^{+0.0202+0.0474}_{-0.0250-0.0698}$&${-1.0076}^{+0.0246+0.0753}_{-0.0299-0.1198}$
&${+2.9200}^{+3.0693+4.9697}_{-4.5900-12.162}$&${-1.0020}^{+0.0082+0.0252}_{-0.0092-0.0346}$&${+0.2270}^{+0.0085+0.0176}_{-0.0114-0.0269}$\\[0.35mm]
\end{tabular}
\end{ruledtabular}
\end{table*}
\endgroup
\begingroup
\squeezetable
\begin{table*}[!hbtp]
\centering
\caption{Shows the effects of the reconstructions of ${\rm I}_{\rm Q}(z)$ and $\omega$ on the parameters $\Omega_{DM}$, $q$ and $j$ in determined $z$ ranges.}\label{effects}
\vspace{0.1cm}
\begin{tabular}{| c | c | c | c |}
\hline\hline
 Model&$XCPL(I)/DR(2)$&$XCPL(II)/DR(1)$&$XCPL(I)/XCPL(II)$\\
 \hline\hline
 Effect&Enhancement&Suppression&Enhancement\\
 Parameter&on $\Omega_{DM}$&on $\Omega_{DM}$&on $q$\\
 \cline{2-3}
 \cline{3-4} 
\hline\hline
$z$&$(+\infty\rightarrow+3.98)$&$(+\infty\rightarrow+3.98)$&$(+0.1\rightarrow-0.7)$\\[0.35mm]
$\omega$&$--$&$--$&$(-0.998\rightarrow-1.159)$\\[0.35mm]
${\rm I}_{Q}(I_{-})$&$--$&$(-0.01\rightarrow-0.1)$&$(-1.384\rightarrow-6.24)\times10^{-3}$\\[0.35mm]
${\rm I}_{Q}(I_{+})$&$(+0.01\rightarrow+0.1)$&$--$&$(+1.504\rightarrow13.976)\times10^{-3}$\\[0.35mm]
$\Omega_{DM}$&$(+0.80\rightarrow +0.85)$&$(+0.80\rightarrow +0.70)$&$(0.178\rightarrow0)$\\[0.35mm]
$q$&$--$&$--$&$(-0.7\rightarrow-1.3)$\\[0.35mm]
\hline
\end{tabular}
\centering\begin{tabular}{| c | c | c | c | c | c | c |}
\hline\hline
 Model&$DR(1)$&$XCPL(I)$&$XCPL(II)$&$DR(1)$&$DR(2)$&$XCPL(I)$\\
 \hline\hline
 Effect&Enhancement&Enhancement&Enhancement&Suppression&Suppression&Suppression\\
 Parameter&on $j$&on $j$&on $j$&on $j$&on $j$&on $j$\\
 \cline{2-3}
 \cline{3-4}
\hline\hline
$z$&$(+8\rightarrow0.295)$&$(+1.723\rightarrow0.045)$&$(+8\rightarrow0.045)$&$(+0.295\rightarrow-1.0)$&$(+8.0\rightarrow-1.0)$&$(+8\rightarrow+1.723)$\\[0.35mm]
$\omega$&$(-1.001\rightarrow-1.002)$&$(-0.980\rightarrow-1.002)$&$(-0.982\rightarrow-1.002)$&$(-1.0019\rightarrow-1.0021)$&
$(-1.0013\rightarrow-1.0021)$&$(-0.9449\rightarrow-0.9619)$\\[0.35mm]
${\rm I}_{Q}\times10^{+3}$&$(-76.65\rightarrow-2.524)$&$(+17.217\rightarrow+0.968)$&$(-153.064\rightarrow-0.320)$&$(-2.524\rightarrow+9.92)$&
$(+19.51\rightarrow-2.09)$&$(+77.976\rightarrow+17.217)$\\[0.35mm]
$\Omega_{DM}$&$(+0.591\rightarrow+0.346)$&$(+0.496\rightarrow+0.258)$&$(+0.461\rightarrow+0.241)$&$(+0.349\rightarrow+0.051)$&
$(+0.346\rightarrow+0.003)$&$(+0.496\rightarrow+0.388)$\\[0.35mm]
$j$&$(-0.9953\rightarrow-1.0)$&$(-0.9968\rightarrow-1.0698)$&$(-0.9821\rightarrow-1.0698)$&$(-1.0003\rightarrow-1.0098)$&$(-1.0061\rightarrow-1.0098)$&
$(-0.9968\rightarrow-1.0176)$\\[0.35mm]
\hline\hline
\end{tabular}
\end{table*}
\endgroup
\begin{figure*}[!htb]
 \centering\includegraphics[width=15.7cm,height=7.7cm]{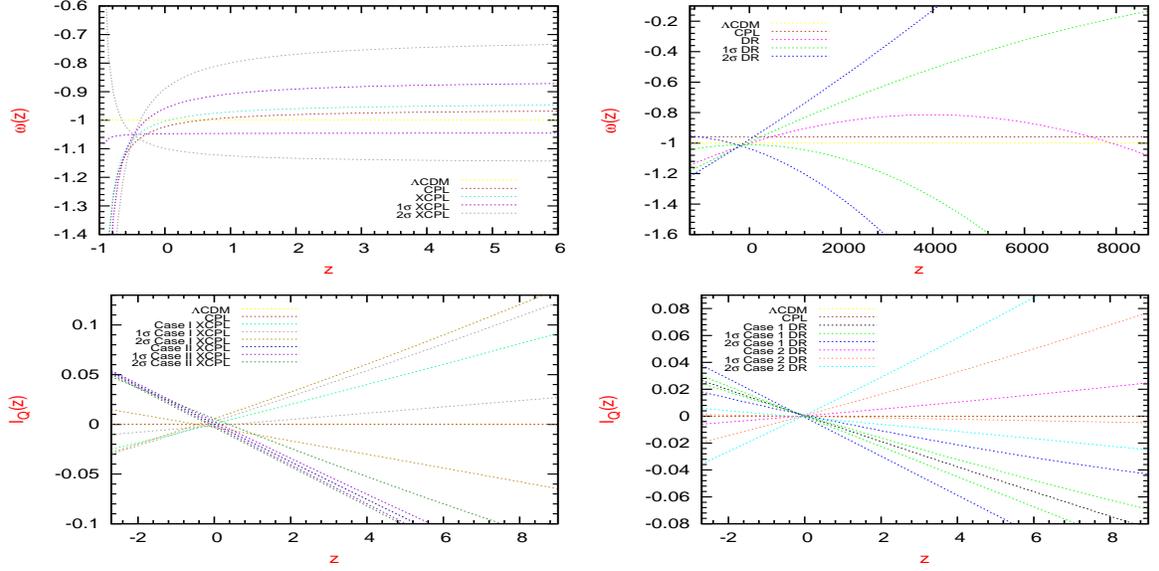}\\
 \caption{(color online) The upper panels display the best reconstructed $\omega(z)$ along $z$, their errors at $1\sigma$ and $2\sigma$ for the coupled models. These results are 
 consistent with the $\Lambda$CDM model predictions. Similarly, The lower panels show the reconstructed evolution of ${\rm I}_{\rm Q}$ and their errors at $1\sigma$ and $2\sigma$ in function of $z$ for 
the coupled models and compared with the predictions of the uncoupled models.}\label{Iq}
\end{figure*}
\begin{figure*}[!htb]
 \centering \includegraphics[width=15.7cm,height=7.7cm]{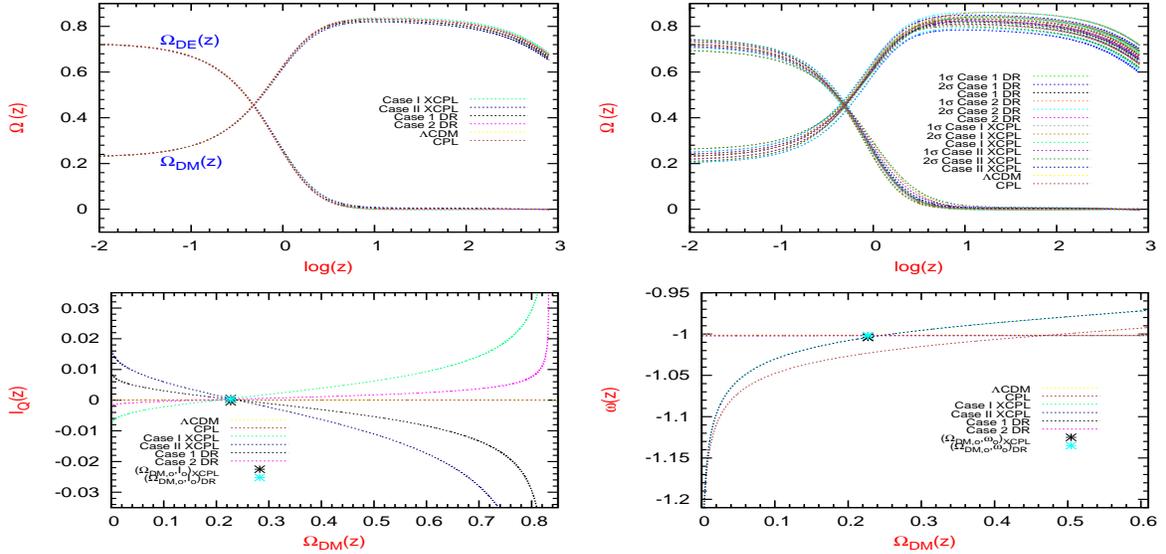}\\
 \caption{(color online) The upper panels shows the best reconstructed energy densities at different $z$ and their errors at $1\sigma$ and $2\sigma$ for the coupled models.
  Here, for $z \geq 3.98$ and ${\rm I}_{+}$, the values of the amplitudes of $\Omega_{DM}$ are amplified, instead, at this same region and for ${\rm I}_{-} <0$, 
  these amplitudes are suppressed with respect to the uncoupled models. The left below panel shows the evolution of ${\rm I}_{\rm Q}$ as function of $\Omega_{DM}$. 
  For $\Omega_{DM}\geq 0.25$, a disminution in the concentration of $\Omega_{DM}$ implies that the amplitude of ${\rm I}_{\rm Q}$ is increased and reduced in the 
  coupled models, instead, for $\Omega_{DM} < 0.25$, it is always increased. Also, the right below panel shows the evolution of $\omega$ as function of $\Omega_{DM}$. 
  Here, a disminution of the values of $\Omega_{DM}$ imply an increasing of the amplitude of $\omega$. Moreover, the ``$o$'' denotes the best fitting parameters in the 
  present (see Table \ref{cosmological_state}).}\label{Omegas}
\end{figure*}
\begin{figure*}[!htb]
 \centering\includegraphics[width=15.7cm,height=7.7cm]{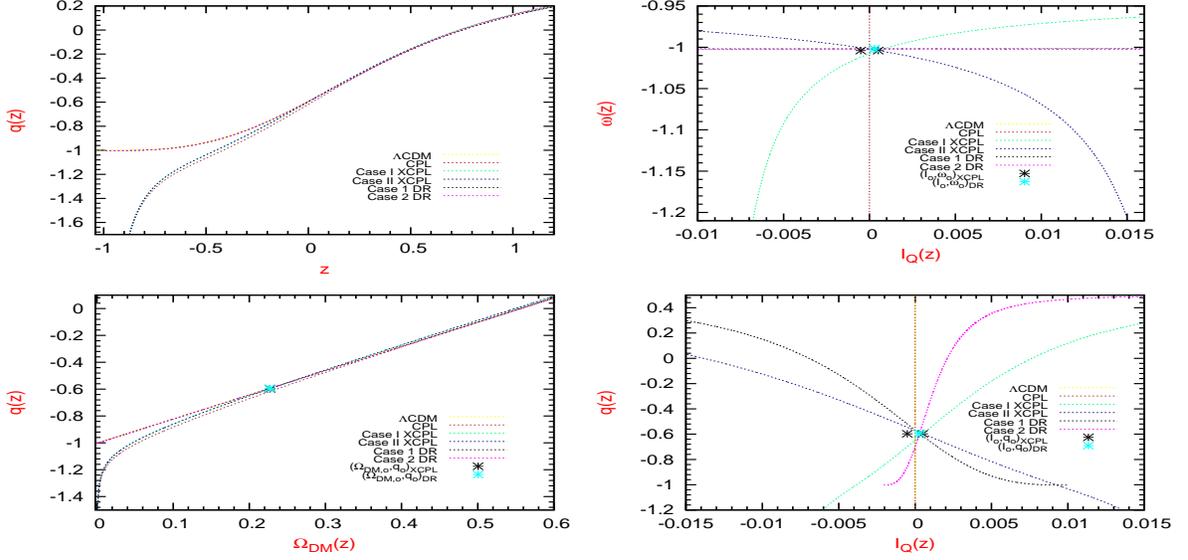}\\
 \caption{(color online) The left above panel shows the evolution of $q$ along $z$.
The right panel displays the evolution of $\omega$ as function of ${\rm I}_{\rm Q}$. Here, the magnitude of $\omega$, 
$\lvert \omega \rvert$ is enhanced from $\lvert-0.95\rvert$ to $\lvert-1.2\rvert$ when the values of ${\rm I}_{\rm Q}$ moves from $0$ to $\lvert0.015\rvert$. 
The left and right below panels, depict the evolution of $q$ as function of $\Omega_{DM}$ and ${\rm I}_{\rm Q}$, respectively. Here, the magnitude of $q$, 
$\lvert q \rvert$, is amplified from $0.1$ to $\lvert-1.4\rvert$ or from $0.4$ to $\lvert-1.2\rvert$, when the values of $\Omega_{DM}$ change from $0.6$ to $0$ or
when the values of ${\rm I}_{\rm Q}(z)$ moves from $0$ to $\lvert0.015\rvert$, respectively. Here, the ``$o$'' denotes the best fitting parameters in the present 
(see Table \ref{cosmological_state}).}\label{geometricalq}
\end{figure*}
\begin{figure*}[!htb]
 \centering\includegraphics[width=15.7cm,height=7.7cm]{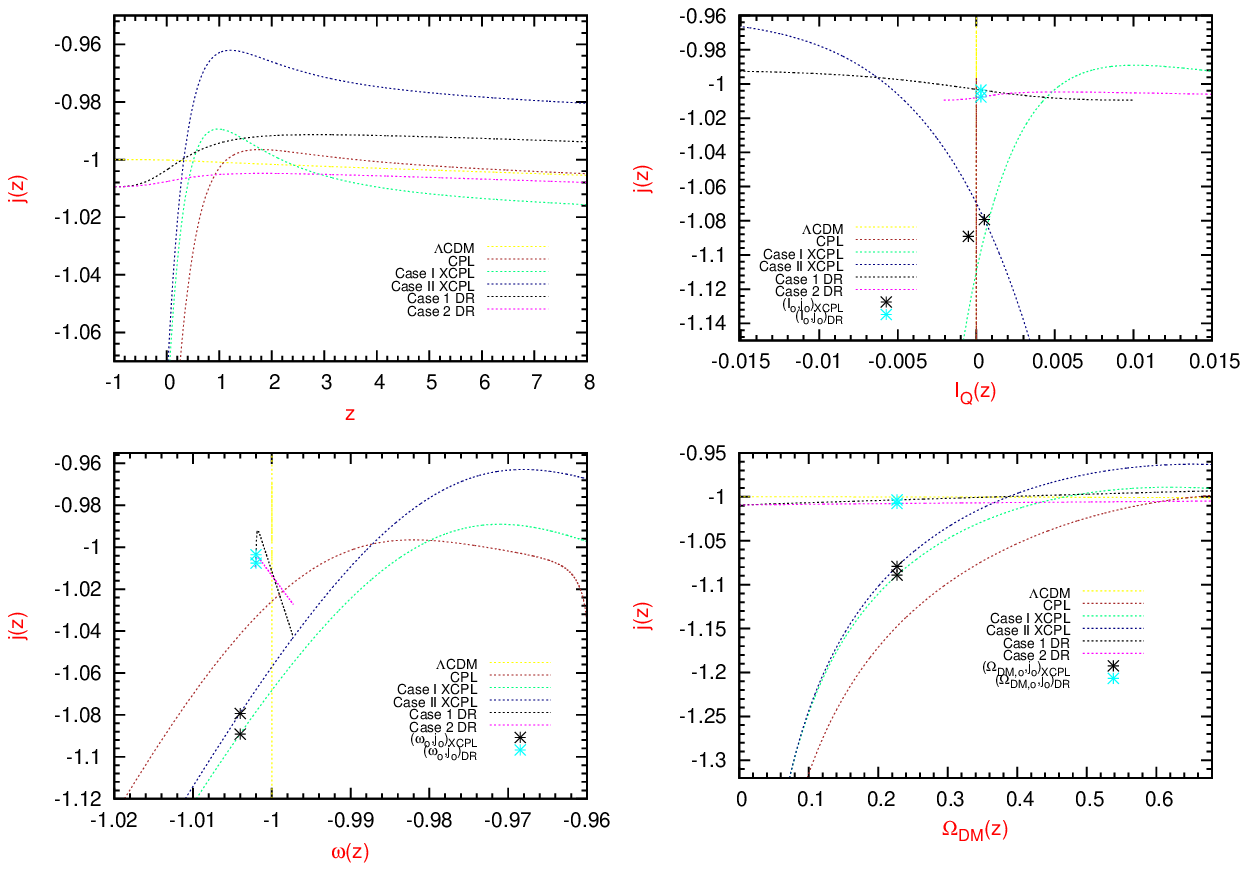}\\
 \caption{(color online) The left above panel shows the evolution of $j$ along $z$. The right panel displays the evolution of $j$ as function of 
 ${\rm I}_{\rm Q}$. Here, the magnitude $\lvert j \rvert$, is amplified from $\lvert-0.96\rvert$ to $\lvert-1.14\rvert$, when the values of  
 ${\rm I}_{\rm Q}(z)$ moves from $0$ to $0.005$ and from $0$ to $-0.001$, respectively. The left and right below panels, depict the evolution of $j$ as function of 
 $\omega$ and $\Omega_{DM}$, respectively. Here, the magnitude $\lvert j \rvert$, is increased from $\lvert-0.96\rvert$ to $\lvert-1.12\rvert$ or 
 from $\lvert-0.95\rvert$ to $\lvert-1.3\rvert$, when the values of $\omega$ moves from $-0.96$ to $-1.02$,  and also, when the values of $\Omega_{DM}$ change from 
 $0.7$ to $0$, respectively. The ``$o$'' denotes the best fitting values of the parameters in the present (see Table \ref{cosmological_state}).}\label{geometricalj}
\end{figure*}
\section{Results}
In this section, we present the results of the fitting on the models listed in Table \ref{Bestfits}, using the Union $2.1$ SNIa 
data set, the BAO data set, the CMB data from WMAP $7$, the H data set and the priors described in Table \ref{Priors}.
Likewise, for the uncoupled $\Lambda$CDM and CPL models, the corresponding free parameters to be estimated are: $\mathbf{X}=(\Omega_{DM,0},H_{0})$ and 
$\mathbf{X}=(\Omega_{DM,0},H_{0},\omega_{o},\omega_{1})$. Meanwhile, for the coupled XCPL and DR models the free parameters are: 
$\mathbf{X}=(\Omega_{DM,0},H_{0},\omega_{o},\omega_{1},\lambda_{0},\lambda_{1},\lambda_{2})$, 
$\mathbf{X}=(\Omega_{DM,0},H_{0},\omega_{o},\omega_{1},\omega_{2},\lambda_{0},\lambda_{1},\lambda_{2})$, respectively.\\ 
In each model, the function $\tilde{\chi}^2_{min}$, the one-dimension probability contours, the best fitting parameters, and their errors at 
$1\sigma$ and $2\sigma$ were computed, by using the Bayesian statistic method, as are shown in Figures \ref{LCDM-CPL_CI}, \ref{XCPL_CI} and \ref{DR_CI}, respectively.\\
The values of the functions $q$, $j$, ${\rm I}_{\rm Q}$, $\omega$ and $\Omega_{DM}$ evalua-ted in $z=0$ (today) are denoted as $q_{o}$, $j_{o}$, ${\rm I}_{o}$, 
$\omega_{o}$ and $\Omega_{DM,0}$, respectively, and are presented in Table \ref{cosmological_state}.\\
Furthermore, the coupled models show that the $\omega=-1$ crossing feature is more favored by the reconstructed $\omega$ in the DR model with
two crossings in the past ($z=21.11$ and $z=7895$) than that found in the XCPL model with only one crossing in a recent epoch ($z=0.07$) \cite{Nesseris2007}. 
These crossing points obtained from the best reconstructed $\omega$ are illustrated in Figure \ref{Iq}.\\
Let us now see Figure \ref{Iq}, within the coupled models have considered that ${\rm I}_{+}$ denotes an energy transfer from $DE$ to $DM$, instead, ${\rm I}_{-}$ 
denotes an energy transfer from $DM$ to $DE$. A change of sign on the best reconstructed ${\rm I}_{\rm Q}$ is linked to the crossing of the noncoupling line 
${\rm I}_{\rm Q}(z)=0$. In this regard, within the coupled models have found a change from ${\rm I}_{+}$ in the past to ${\rm I}_{-}$ in the present and vice versa. 
According to this Figure and Table \ref{cosmological_state} note that a non-negligible value of ${\rm I}_{0}$ at $1\sigma$ error has been found in the coupled 
models, and whose order of magnitude is in agreement with the results obtained in \cite{abramo1, Cai-Su, abramo2, cao2011, LiZhang2011, Cueva-Nucamendi2012}.\\
Due to the two minimums obtained in each coupled model (see Table \ref{Bestfits}), then two different possibilities to 
reconstruct ${\rm I}_{\rm Q}$ have been found here. Therefore, for any $z$ range, an interesting mixture of ${\rm I}_{+}$ and ${\rm I}_{-}$ may be des-cribed there.\\  
In what follows we compared both the results of the XCPL model with those of the CPL model, and also, the predictions of the DR model with the corresponding of the 
$\Lambda$CDM model. Otherwise, according to the results presented in Figures \ref{Iq} and \ref{Omegas}. For $z \geq 3.98$ the values of the amplitudes of $\Omega_{DM}$ in 
the coupled models are slightly modified by the values of ${\rm I}_{\rm Q}$ (${\rm I}_{+}$ or ${\rm I}_{-}$) when they are compared with the uncoupled 
models. For ${\rm I}_{+} >0$, the amplitudes of $\Omega_{DM}$ are amplified, instead, for ${\rm I}_{-} <0$, these amplitudes are suppressed. 
These results coincide with those found in \cite{cabral2009}. In addition, in any another $z$ region, these effects are not possible. They are described in 
Table \ref{effects} (left and center above table), in where the $z$ ranges and the coupled models are indicated. Likewise, we have also noted that, the shape of 
$\Omega_{DE}$ and the valu-es of its amplitudes are not significantly affected by the reconstructions of ${\rm I}_{\rm Q}$ and $\omega$ with respect to uncoupled models. 
Furthermore, we also confirm that the coincidence problem is alleviated in these coupled models, but they may not solve it. The below panels in Figure \ref{Omegas}, 
as well as, the right above panel and below panels in Figures \ref{geometricalq} and \ref{geometricalj}, the constraints at $1\sigma$ and $2\sigma$ 
on $\Omega_{DM}$, $\Omega_{DE}$, ${\rm I}_{\rm Q}$, $\omega$, $q$ and $j$ have been omitted to obtain a better visuali-zation of these effects.\\
From Figure \ref{geometricalq} (left above panel) our coupled models have predicted that a transition from a deceleration era at early times to an acceleration era at late 
times has been reached by the universe and the redshift for this change, $z_{T}$, is found to be $z_{T}\sim 0.75$.\\ 
Then, at $z\rightarrow -1$ the DR model cannot determine, if the big-rip \cite{Caldwell2003} may or may not occur in the universe, instead, for the XCPL model the universe 
will finish in a big-rip. These results coincide with those obtained in \cite{Li-Ma}. However, from Figure \ref{geometricalq} (left above panel) and Table \ref{effects} 
(right above table) our results in the XCPL model have revealed that in determined $z$ intervals
the amplitude of $q$ is slightly enhanced due to the increasing of the magnitudes of ${\rm I}_{\rm Q}$ and $\omega$ (see right above panel and below panels in 
Figure \ref{geometricalq}) with respect to that of the CPL model. Such a effect was realized when $\Omega_{DM}$ becomes less concentrated at $\log(z)\in[-2,+0.3]$ 
(see left panel in Figure \ref{Omegas}). It was the epoch of the $DE$ dominance, from which, the universe was led to an accelerated expansion.\\Regarding Figure \ref{geometricalj} (left above panel), 
we have found qualitatively different asymptotic values in the near future $(z\rightarrow -1)$, but similar asymptotic values in the past ($z\gg1$) are exhibited by the best 
reconstructed $j$. Now let us analyze this Figure (right upper panel and lower panels) and Table \ref{effects} (below tables), from which, note that the magnitudes of 
${\rm I}_{\rm Q}$ and $\omega$ have imprinted new physical effects on the amplitudes of the parameter $j$. Within the coupled 
models the amplitudes of $j$ were progressively increased or reduced in determined $z$ regions, with respect to those of the uncoupled models for increasing of the 
magnitudes of ${\rm I}_{\rm Q}$ and $\omega$. Furthermore, the expansion of the universe was modified by the consequent diminution of $\Omega_{DM}$. Therefore, we have 
shown that las magnitudes of ${\rm I}_{\rm Q}$ and $\omega$ are strongly related with the magnitudes of $\Omega_{DM}$, $q$ and $j$, respectively, 
as are seen in Figures \ref{Omegas}, \ref{geometricalq} and \ref{geometricalj}.\\
We now compare our results with those obtained by other researchers. In \cite{Qi2009} several parametrizations for $\omega$ were proposed, and also, a series 
for $\omega$ in \cite{Rubano2012} (see Tables $1$ and $2$), was investigated, and then, compared it with a expansion of the scale factor up the fifth order. 
The results of \cite{Qi2009} and \cite{Rubano2012} are comparable with our results at $1\sigma$ error (see Table \ref{cosmological_state}).
Furthemore, in \cite{Viel2010} and \cite{Gruber2012} the values of $q_{o}$ and $j_{o}$ were estimated from a series, in where, new variables were defined 
to avoid the problem of divergence, for example, see both Table $2$ in \cite{Viel2010} and Tables I, II and III in \cite{Gruber2012}, respectively. 
These results are compatible at $1\sigma$ error with those found in Table \ref{cosmological_state}.
Otherwise, constraints on $q_{o}$ and $j_{o}$ were established by the authors in \cite{Sendra2013} (see Table $2$, series $2$D and $3$D) from a general 
expression of the BAO modes, and also, employing a Taylor expansion and Pad\'e approximations in \cite{Gruber2014} (see Table I, fourth column, fit 3). 
The results obtained in \cite{Sendra2013} and \cite{Gruber2014} at $1\sigma$ are also consistent with those presented in Table \ref{cosmological_state}.
\section{Conclusions}\label{SectionConclusions}
Now we summarize our main results:\\
$\bullet$ An analysis combined of data was performed to break the degeneracy among the cosmological parameters of our models, allow us to obtain constraints more 
stringent on them. In particular, for the XCPL and DR models, the allowed region of its parameters was significantly reduced by the inclusion of the CMB data,
compared with studies of models without the CMB data \cite{Cueva-Nucamendi2012, He-Wang(2011)}. This implies that higher redshift may be able to discriminate between
these models.\\
$\bullet$ In the DR model, a novel reconstruction for $\omega$ was proposed whose best fitted value is closed to $-1$, and has the property of avoiding divergences in a 
distant future $z\rightarrow -1$. This result is consistent with the value predicted by the $\Lambda$CDM model at $1\sigma$ error. Lisewise, within this scenario,
a finite value for $\omega$ has been obtained from the past to the future, mainly, the asymptotic values are: $\omega(z)=\omega_{2}z^{2}$ for $z\gg1$, $\omega(z)\approx \omega_{o}$ for $z\ll 1$ and 
$\omega(z)\approx \omega_{o}$ for $z\rightarrow -1$. Therefore, a better physical description of the dynamical evolution of $DE$ is performed by the DR model, which 
should be used to explore the properties of $DE$.\\
$\bullet$ Currently, a phase of accelerated expansion is the situa-tion revealed by all our models about the universe. The big-rip problem is not forecasted by the 
DR model, and hence, this scenario should be considered to study the ultimate destinity of the universe. Likewise, from the coupled models found that the values of the
amplitudes of the parameter $q$ are not significantly affected neither by values of ${\rm I}_{\rm Q}$ nor by the values of $\omega$ (see left above panel in Figure 
\ref{geometricalq}).\\
$\bullet$ The values of the amplitudes of $\Omega_{DE}$ (see upper panels in Figure \ref{Omegas}) are not significantly modified by the recons-tructions of ${\rm I}_{\rm Q}$ and $\omega$, respectively, nevertheless, 
they are definitely positive. This requirement implies that $\omega$ must be always negative in all the cosmic stages of the universe (see upper panels in Figure \ref{Iq}).\\   
$\bullet$ The coupled models are strongly favored by the observational data having a preference for $j_{0}<-1$, and hence, they represent a slight deviation from 
the value predicted by the $\Lambda$CDM model (see Table \ref{cosmological_state}).\\ 
$\bullet$ The behaviours qualitatively presented here show that the graph of $j$ has more possibility in discriminating the different coupled $DE$ models, 
and therefore, $j$ could be used to distinguish them (see left upper panel in Figure \ref{geometricalj}).\\    
$\bullet$ The physical effects generated by the magnitudes of ${\rm I}_{\rm Q}$ and $\omega$ on the cosmological parameters could be understood, so:
An energy transfer from $DE$ to $DM$ or vice versa inserts energy into one of the fluids, and determines an increase of the energy density on one of them,
which increases the Hubble parameter inducing a slight expansion of the universe, and to recover equilibrium of the system ${\rm I}_{\rm Q}$ and $\omega$ leads to an 
enhancement or suppression on the amplitudes and shapes of $\Omega_{DM}$, $q$ and $j$ with respect to uncoupled models, in determined redshift ranges.\\
In a forthcoming paper we will extend our study by applying cosmological perturbation theory on the coupled models, using data of linear matter power spectrum, 
weak lensing potential, integrated Sach-Wolfe, growth rate, and other. They will allow us to calculate of how the magnitudes of ${\rm I}_{\rm Q}$ and $\omega$ operate on the 
amplitudes of $\Omega_{DM}$, $q$ and $j$, respectively. From which, we may conclude if our DR model can emerge as an alternative to the $\Lambda$CDM model. 
This will be the purpose of our future work.
\appendix*
\section{Integrals $I_n(z)$ and $\tilde{I_{n}}(\tilde{x})$}\label{integralIn}
\begin{widetext}
\begin{eqnarray}
I_{0}(z)&=&\frac{2}{z_{max}}\biggl[\ln\biggl(1+z\biggr)\biggr]\;,\\
I_{1}(z)&=&\frac{2}{z_{max}}\biggl[\frac{2z}{z_{max}}-\frac{(2+z_{max})}{z_{max}}\ln\biggl(1+z\biggr)\biggr]\;,\\
I_{2}(z)&=&\frac{2}{z_{max}}\biggl[\frac{4 z}{z_{max}}\left(\frac{z}{z_{max}}-\frac{2}{z_{max}}-2\right)+
\left(1+\frac{6.828427}{z_{max}}\right)\left(1+\frac{1.171572}{z_{max}}\right)\ln\biggl(1+z\biggr)\biggr]\;,\\
\tilde{I_{0}}(\tilde{x})&=&\frac{2}{z_{max}}\biggl[\ln{\biggl(1+0.5\;z_{max}(1+\tilde{x})\biggr)}\biggr]\;,\\
\tilde{I_{1}}(\tilde{x})&=&{\frac{2}{z_{max}}}\biggl[\biggl(1+\tilde{x}\biggr)-\frac{(2+z_{max})}{z_{max}}\ln{\biggl(1+0.5\;z_{max}(1+\tilde{x})\biggr)}\biggr]\;,\\
\tilde{I_{2}}(\tilde{x})&=&{\frac{2}{z_{max}}}\biggl[\biggl(1+\tilde{x}\biggr)\biggl(\tilde{x}-\frac{4}{z_{max}}-3\biggr)+\biggl(1+\frac{6.828427}{z_{max}}\biggr)
\biggl(1+\frac{1.171572}{z_{max}}\biggr)\ln{\biggl(1+0.5\;z_{max}(1+\tilde{x})\biggr)}\biggr]\;.
\end{eqnarray}
\end{widetext}
\begin{acknowledgments}
The author is grateful to Prof. F. Astorga for his academic support and fruitful discussions in the early stages of this research, thank Prof. O. Sarbach and 
Prof. L. Ure\~na for useful discussions and comments, respectively. This work was in beginning supported by the IFM-UMSNH.
\end{acknowledgments}
 

\begin{thebibliography}{78}
\begin{scriptsize}
\bibitem{Riess1998}
A.G. Riess et al., \textit{Astron. J.} \textbf{116} (1998) 1009; S. Perlmutter et al; \textit{Astrophys. J.} \textbf{517} (1999) 565; 
J.L. Tonry et al., \textit{Astrophys. J.} \textbf{659} (2007) 98; T.M. Davis et al., \textit{Astrophys. J.} \textbf{666} (2007) 716; 
M. Kowalski et al., \textit{Astrophys. J.} \textbf{686} (2008) 749.
\bibitem{AmanullahUnion22010}
R. Amanullah et al., \textit{Astrophys. J.} \textbf{716} (2010) 712.
\bibitem{nesseris1}
S. Nesseris and L. Perivolaropoulos, \textit{J. Cosmol. Astropart. Phys.} \textbf{0702} (2007) 025.
\bibitem{Suzuki2012}
N. Suzuki et al., \textit{Astrophys. J.} \textbf{85} (2012) 746.
\bibitem{Eisenstein1998}
D. J. Eisenstein, W. Hu, \textit{Astrophys. J.} \textbf{496} (1998) 605.
\bibitem{SDSS}
K. Abazajian et al., \textit{Astron. J.} \textbf{126} (2003) 2081; \textit{Astron. J.} \textbf{128} (2004) 502; \textit{Astron. J.} \textbf{129} (2005) 1755;
\textit{Astrophys. J. Suppl.} \textbf{182} (2009) 543; M. Tegmark et al., \textit{Phys. Rev.} \textbf{D 69} (2004) 103501;
\textit{Phys. Rev.} \textbf{D 74} (2006) 123507; D. J. Eisenstein et al., \textit{Astrophys. J.} \textbf{633} (2005) 560.
\bibitem{ReidSDSS}
B. A. Reid et al., \textit{Mon. Not. Roy. Astron. Soc.} \textbf{401} (2010) 2148; \textit{Mon. Not. Roy. Astron. Soc.} \textbf{404} (2010) 60.
\bibitem{Beutler2011}
F. Beutler et al., \textit{arxiv: 1106.3366v1} \textbf{06} (2011) 16.  
\bibitem{Blake2011}
C. Blake et al., \textit{Mon. Not. Roy. Astron. Soc.} \textbf{418} (2011) 1707-1724. 
\bibitem{Hu-Sugiyama1996}
W. Hu and N. Sugiyama, \textit{Astrophys. J.} \textbf{471} (1996) 542.
\bibitem{Bond-Tegmark1997}
J. R. Bond, G. Efstathiou and M. Tegmark, \textit{Mon. Not. Roy. Astron. Soc.} \textbf{291} (1997) L33.
\bibitem{WMAP}
C.L. Bennett et al., \textit{Astrophys. J. Suppl.} \textbf{148} (2003) 1; D.N. Spergel et al., \textit{Astrophys. J. Suppl.} \textbf{148} (2003) 175;
\textit{Astrophys. J. Suppl.} \textbf{170} (2007) 377; G. Hinshaw et al., \textit{Astrophys. J. Suppl.} \textbf{180} (2009) 225;
E. Komatsu et al., \textit{Astrophys. J. Suppl.} \textbf{180} (2009) 330.
\bibitem{Komatsu2011}
E. Komatsu et al., \textit{Astrophys. J. Suppl.} \textbf{192} (2011) 18.
\bibitem{Jimenez2002}
R. Jimenez and A. Loeb, \textit{Astrophys. J.} \textbf{573} (2002) 37.
\bibitem{Jimenez2003}
R. Jimenez,L. Verde, T. Treu and D.Stern, \textit{Astrophys. J.} \textbf{593} (2003) 622.
\bibitem{Simon2005}
J. Simon, L. Verde and R. Jimenez, \textit{Phys. Rev.} \textbf{D 71} (2005) 123001.
\bibitem{Hubble2009-Stern2010}
A. G. Riess et al., \textit{Astrophys. J.} \textbf{699} (2009) 539;
D. Stern, R. Jimenez, L. Verde, M. Kamionkowski, S. A. Stanford, \textit{Astrophys. J. Suppl.} \textbf{188} (2010) 280; 
\textit{J. Cosmol. Astropart. Phys.} \textbf{02} (2010) 8.
\bibitem{Gaztanaga}
E. Gaztanaga, A. Cabre and L. Hui, \textit{Mon. Not. Roy. Astron. Soc.} \textbf{399} (2009) 1663.
\bibitem{Peebles1988}
P. J. E. Peebles and B. Ratra, \textit{Astrophys. J.} \textbf{325} (1988) L17.
\bibitem{Peebles2003}
P. J. E. Peebles and B. Ratra, \textit{Rev. Mod. Phys.} \textbf{75} (2003) 559.
\bibitem{Sahni2004}
V. Sahni, \textit{Lect. Notes Phys.} \textbf{653} (2004) 141.
\bibitem{Copeland2006}      
E. J. Copeland, M. Sami and S. Tsujikawa, \textit{Int. J. Mod. Phys.} \textbf{D 15} (2006) 1753.
\bibitem{Weinberg1989}
S. Weinberg, \textit{Rev. Mod. Phys.} \textbf{61} (1989) 1.
\bibitem{Sahni2000}
V. Sahni and A. A. Starobinsky, \textit{Int. J. Mod. Phys.} \textbf{D 9} (2000) 373.
\bibitem{Seljak2005} 
U. Seljak et al., \textit{Phys. Rev. D} \textbf{71} (2005) 103515. 
\bibitem{Rozo2010}
E. Rozo et al., \textit{Astrophys. J.} \textbf{708} (2010) 645.
\bibitem{Caldwell2002}
R. R. Caldwell, \textit{Phys.Lett. B} \textbf{545} (2002) 23;
S. Nojiri and S. D. Odintsov, \textit{Phys. Lett. B} \textbf{562} (2003) 147;
R. Gannouji, D. Polarski, A. Ranquest and A. A. Starobinsky, \textit{J. Cosmol. Astropart. Phys.} \textbf{09} (2006) 016;
X. Cheng, Y. Gong and E. N. Saridakis, \textit{J. Cosmol. Astropart. Phys.} \textbf{04} (2009) 001.
\bibitem{Feng2005}
E. Elizalde, S. Nojiri, and S. D. Odintsov \textit{Phys. Rev. D} \textbf{70} (2004) 043539;
Z. K. Guo, Y. S. Piao, X. M. Zang and Y. Z. Zhang, \textit{Phys. Lett. B} \textbf{608} (2005) 177.
\bibitem{Ratra1988}
B. Ratra, and P. J. E. Peebles, \textit{Phys. Rev. D} \textbf{37} (1988) 3406;
K. Coble, S. Dodelson, and J. A. Frieman, \textit{Phys. Rev. D} \textbf{55} (1997) 1851;
R. R. Caldwell, R. Dave, and P. J. Steinhardt, \textit{Phys. Rev. Lett.} \textbf{80} (1998) 1582.
\bibitem{Picon-Chiba}
C. Armendariz-Picon, V. Mukhanov, P. J. Steinhardt, \textit{Phys. Rev. Lett.}\textbf{85} (2000) 4438, \textit{Phys. Rev. D} \textbf{63} (2001) 103510;
T. Chiba, T. Okabe, M. Yamaguchi, \textit{Phys. Rev. D} \textbf{62} (2000) 023511.
\bibitem{Pasquier-Harko}
A. Y. Kamenshchik, U. Moschella and V. Pasquier, \textit{Phys. Lett. B} \textbf{511} (2001) 265; 
M. C. Bento, O. Bertolami and A. A. Sen, \textit{Phys. Rev. D} \textbf{66} (2002) 043507;
M. K. Mak and T. Harko, \textit{Phys. Rev. D} \textbf{71} (2005) 104022;
\bibitem{Garousi-Sami}
M. R. Garousi, M. Sami, S. Tsujikawa, \textit{Phys. Lett. B} \textbf{606} (2005) 1;
M. R. Garousi, M. Sami, and S. Tsujikawa, \textit{Phys. Rev. D} \textbf{71} (2005) 083005.
\bibitem{Cooray1999}
A. R. Cooray and D. Huterer, \textit{Astrophys. J.} \textbf{513} (1999) L95.
\bibitem{Chevallier-Linder}
M. Chevallier, D. Polarski, \textit{Int. J. Mod. Phys. D}\textbf{10} (2001) 213;
E. V. Linder, \textit{Phys. Rev. Lett.} \textbf{90} (2003) 091301  
\bibitem{Tegmark2004}
J. Barboza, E. M. and J. Alcaniz, \textit{Phys. Lett. B} \textbf{666} (2008) 415.
\bibitem{Barboza2009}
E. M. Barboza Jr. et al; \textit{Phys. Rev. D.} \textbf{80} (2009) 043521.
\bibitem{Wu2010}
Q. J. Zhang and Y. L. Wu, \textit{J. Cosmol. Astropart. Phys.} \textbf{08} (2010) 038.
\bibitem{Li-Ma}
H. Li and X. Zhang, \textit{Phys. Lett. B} \textbf{703} (2011) 119;
J. Z. Ma and X. Zhang, \textit{Phys. Lett. B} \textbf{699} (2011) 233. 
\bibitem{Holsclaws}
T. Holsclaw et al, \textit{Phys. Rev. Lett.} \textbf{105} (2010) 241302;
T. Holsclaw et al, \textit{Phys. Rev. D.} \textbf{84} (2011) 083501. 
\bibitem{Daly2003}
R. A. Daly and S. Djorgovski, \textit{Astrophys. J.} \textbf{597} (2003) 009.
\bibitem{Huterer2005}
D. Huterer and A. Cooray, \textit{Phys. Rev. D.} \textbf{71} (2005) 023506. 
\bibitem{Alam2006}
A. Shafieloo, U. Alam, V. Sahni and A. A. Starobinsky, \textit{Mon. Not. R. Astron. Soc.} \textbf{366} (2006) 1081.
\bibitem{Hojjati2010}
A. Hojjati, L. Pogosian and G. B. Zhao, \textit{J. Cosmol. Astropart. Phys.} \textbf{04} (2010) 007.
\bibitem{Olivier2012}
O. Sarbach and M. Tiglio, \textit{Liv. Rev. Rel.} \textbf{15} (2012) [gr-qc/1203.6443v1]. 
\bibitem{Martinez2008}
E. F. Martinez and L. Verde, \textit{J. Cosmol. Astropart. Phys.} \textbf{08} (2008) 023.
\bibitem{Turner1983}
M. S. Turner, \textit{Phys. Rev. D} \textbf{28} (1983) 1243.
\bibitem{Malik2003}
K. A. Malik, D. Wands, and C. Ungarelli, \textit{Phys. Rev. D} \textbf{67} (2003) 063516.
\bibitem{Cen2003}
R. Cen, \textit{Astrophys. J.} \textbf{546} (2001) L77;
M. Oguri, K. Takahashi, H. Ohno and K. Kotake, \textit{Astrophys. J.} \textbf{597} (2003) 645. 
\bibitem{Guo2007}
Z. K. Guo, N. Ohta, and S. Tsujikawa, \textit{Phys. Rev. D}\textbf{76} (2007) 023508.   
\bibitem{Bohmer2008}
C. G. Bohmer, G. Caldera-Cabral, R. Lazkoz, and R. Maartens, \textit{Phys. Rev. D} \textbf{78} (2008) 023505.
\bibitem{valiviita2008}
J. Valiviita, E. Majerotto and R. Maartens, \textit{J. Cosmol. Astropart. Phys.} \textbf{07} (2008) 020.
\bibitem{campo2009}
S. Campo, R. Herrera and D. Pavon, \textit{J. Cosmol. Astropart. Phys.} \textbf{01} (2009) 020.
\bibitem{cabral2009}
G. Caldera-Cabral, R. Maartens and B. M. Schaefer, \textit{J. Cosmol. Astropart. Phys.} \textbf{07} (2009) 027.
\bibitem{chimento2010}
L. P. Chimento, \textit{Phys. Rev.}, \textbf{D81} (2010) 043525.
\bibitem{abramo1}
E. Abdalla, L. R. Abramo and J. C. C. de Souza, \textit{Phys. Rev.} \textbf{D 82} (2010) 023508.
\bibitem{Cai-Su}
R. G. Cai and Q. Su, \textit{Phys. Rev.} \textbf{D 81} (2010) 103514.
\bibitem{abramo2}
J. H. He, B. Wang, and E. Abdalla, \textit{Phys. Rev.} \textbf{D 83} (2011) 063515.
\bibitem{cao2011}
S. Cao, N. Liang and Z. H. Zhu, \textit{astro-ph.CO/1105.6274}.
\bibitem{LiZhang2011}
Y. H. Li and X. Zhang, \textit{Eur. Phys. J.} \textbf{C 71} (2011) 1700.
\bibitem{Zimdahl2005}
W. Zimdahl, \textit{Int. J. Mod. Phys. D} \textbf{14} (2005) 2319.
\bibitem{Das2006}
S. Das, P. S. Corasaniti, and J. Khoury, \textit{Phys. Rev. D} \textbf{73} (2006) 083509.
\bibitem{Huey2006}
G. Huey and B. D. Wandelt, \textit{Phys. Rev. D} \textbf{74} (2006) 023519.
\bibitem{Wang2007}
B. Wang, J. Zang, C. Y. Lin, E. Abdalla, and S. Micheletti, \textit{Nucl. Phys. B} \textbf{778} (2007) 69.
\bibitem{Cueva-Nucamendi2012}
F. Cueva Solano and U. Nucamendi, \textit{J. Cosmol. Astropart. Phys.} \textbf{04} (2012) 011;
F. Cueva Solano and U. Nucamendi, \textit{arXiv: 1207.0250} \textbf{07} (2012) 02.
\bibitem{Qi2009}
F. Y. Wang, Z. G. Dai, and Shi Qi, \textit{Astronomy $\&$ Astrophys.} \textbf{507} (2009) 53-59.
\bibitem{Rubano2012}
M. Demianski, E. Piedipalumbo, C. Rubano, P. Scudellaro, \textit{Mon. Not. R. Astron. Soc.} \textbf{426} (2012) 1396-1415.
\bibitem{Viel2010}
 V. Vitagliano, J. Q. Xia, S. Liberati and M. Viel, \textit{J. Cosmol. Astropart. Phys.} \textbf{03} (2010) 005.
\bibitem{Gruber2012}
A. Aviles, C. Gruber, O. Luongo, H. Quevedo, \textit{Phys. Rev. D.} \textbf{86} (2012) 123516.
\bibitem{Sendra2013}
R. Lazkoz, J. Alcaniz, C. Escamilla-Rivera, V. Salzano, I. Sendra, \textit{J. Cosmol. Astropart. Phys.} \textbf{12} (2013) 005.
\bibitem{Gruber2014}
C. Gruber and O. Luongo, \textit{Phys. Rev. D.} \textbf{89} (2014) 103506.
\bibitem{Koyama2009-Brax2010}
Kazuya Koyama, Roy Maartens, and Yong-Seon Song, \textit{J. Cosmol. Astropart. Phys.} \textbf{10} (2009) 017;
P. Brax, C. van de Bruck, D. F. Mota, N. J. Nunes, and H. A. Winther, \textit{Phys. Rev. D} \textbf{82} (2010) 083503.
\bibitem{Lazkoz2007}
R. Lazkoz and E. Majerotto, \textit{J. Cosmol. Astropart. Phys.} \textbf{07} (2007) 015;
J. Lu, L. Xu, M. Liu and Y. Gui, \textit{Eur. Phys. J. C} \textbf{58} (2008) 311;
L. Samushia and B. Ratra, \textit{Astrophys. J.} \textbf{650} (2006) L5.
\bibitem{Lu2009}
J. B. Lu, Y. X. Gui and L. X. Xu, \textit{Eur. Phys. J. C} \textbf{63} (2009) 349.
\bibitem{Xu2010}
L. X. Xu and J. B. Lu, \textit{J. Cosmol. Astropart. Phys.} \textbf{03} (2010) 025.
\bibitem{Feng2011}
L. Feng and Y. P. Yang, \textit{Astron. Astrophys.} \textbf{11} (2011) 751. 
\bibitem{Nesseris2007}
S. Nesseris and L. Perivolaropoulos, \textit{JCAP} \textbf{01} (2007) 018.
\bibitem{Caldwell2003}
R. R. Caldwell, M. Kamionkowski and N. N. Weinberg, \textit{Phys. Rev. Lett.} \textbf{91} 071301 (2003);
L. P. Chimento and R. Lazkoz, \textit{Mod. Phys. Lett. A} \textbf{19} 2479 (2004).
\bibitem{He-Wang(2011)}
Xiao-Dong Xu, Jian-Hua He, Bin Wang, \textit{Phys. Lett. B} \textbf{701} (2011) 513-519. 
\end{scriptsize}
  \end{thebibliography}
\end{document}